\documentclass[twocolumn,epjc3]{svjour3}          % twocolumn
\journalname{Eur. Phys. J. A}
\RequirePackage{graphicx}
\def\nn{\nonumber\\}    
\def\l{\lambda}    
\def\sl1{\sqrt{\lambda_1}}    
\def\sl2{\sqrt{\lambda_2}}    
\def\sl3{\sqrt{\lambda_3}}    
\def\slm{\sqrt{\lambda_m}}    
\def\slq{\sqrt{\lambda_q}}    
\def\g{\gamma}    
\def\d{\delta}    
\def\G{\Gamma}    
\def\be{\begin{eqnarray}}    
\def\ee{\end{eqnarray}}    
\begin{document}
\title{
Radiative corrections to the lepton current in unpolarized elastic $lp$-interaction
for fixed $Q^2$ and scattering angle 
}
\author{ A. Afanasev\thanksref{e1,addr1}
            \and
     A. Ilyichev\thanksref{e2,addr2,addr3}}

\thankstext{e1}{e-mail: afanas@gwu.edu}
\thankstext{e2}{e-mail: ily@hep.by}

\institute{Department of Physics,
The George Washington University,
Washington, DC 20052 USA\label{addr1}
\and
 Belarusian State University,
220030  Minsk,  Belarus\label{addr2}
\and
Institute for Nuclear Problems,
 Belarusian State University,
220006  Minsk,  Belarus\label{addr3}
}

\date{Received: date / Accepted: date}

\maketitle
\begin{abstract}
The kinematical difference between the description of radiative effects for fixed $Q^2$ vs a fixed scattering angle in the elastic lepton-proton ($lp$)-scattering is discussed.   
The technique of calculation as well as explicit expressions for radiative corrections to the lepton current in unpolarized elastic $lp$-scattering for  these two cases are presented without using an ultrarelativistic approximation. A comparative numerical analysis within kinematic conditions of Jefferson Lab measurements and MUSE experiment in PSI is performed. 
\end{abstract}
\section{Introduction}
The elastic lepton-proton scattering is a recognized tool for investigation 
of the internal proton structure. The observation of the disagreement in $Q^2$-behavior of the proton elastic form factor ratio for unpolarized \cite{Andivahis94,Qattan05}  and polarized \cite{Jones,Gayou} electron scattering, along with the proton radius puzzle coming from the different outcomes of the
measurements in electron-proton systems \cite{CODATA,Sick} and in  the muonic hydrogen
\cite{Pohl} -- all of these require understanding of underlying QED processes that may lead to systematic uncertainties at a per cent level. Moreover, the results of the
recent experiment PRAD \cite{PRAD} was in agreement with muonium spectroscopy experiment that contradicted the previous electron-proton scattering data. This unexpected result motivates new efforts for the theoretical and experimental investigations.

One of the important and essential tools for the investigation of the electromagnetic properties of the proton is an experimental program with high duty-cycle positron beams at JLab \cite{JLabpos}. This program with the electron beams allows to estimate the electromagnetic form factors of the proton separately as well as to measure a change asymmetry that appears at the lowest order as an interference of the matrix elements with one- and two-photon exchanges.

Together with widely discussed two-photon exchange \cite{Afanasev_review}, the important
source of uncertainties for both lepton and anti-lepton scattering is from the real photon emission accompanying any process with the charge particle scattering, as well as the additional virtual particle contributions. Due to smallness of muon beam momentum at MUSE experiment in PSI \cite{MUSE}, as well as scattering by extremely small angles  in  PRAD-II experiment at Jefferson Lab \cite{PRAD2}, all calculations have to be performed beyond the ultrarelativistic approximation, $i.e.$ retaining lepton's mass during the entire calculation. While for purely elastic scattering at a given beam energy the four-momentum transfer $Q^2$ is in one-to-one correspondence with a lepton scattering angle, this is not the case for radiative events. It is therefore of critical importance to understand the role of QED radiative corrections (RC) in different kinematic scenarios: fixed momentum transfer $Q^2$ vs fixed scattering angle of the detected lepton (as done in MUSE  \cite{MUSE}  or in high-resolution spectrometers with small angular acceptance used in some of Jefferson Lab experiments).

It should be noted that rather often for estimation of the similar corrections to the exclusive process the additional particle contributions are calculated exactly or within ultrarelativistic approximation (with respect to lepton’s mass) while the real photon emission is considered within the soft photon approximation. Particularly in the papers \cite{Kaiser} and \cite{Vanderhaeghen} 
for M\"oller and virtual Compton scattering processes, respectively,
the virtual QED corrections have been calculated beyond
the ultrarelativistic limit but only the soft part of the real photon emission was  taken into account. 

Mo and Tsai first developed a systemic approach to calculate RC with hard photon emission in elastic and inelastic electron-proton scattering \cite{MoTsai1969}. One limitation in their calculations was the approximate way to consider the soft-photon contribution, as a result, their final expressions depend on an artificial parameter $ \Delta$ that was introduced  to separate the photon momentum phase space into soft and hard parts.

Here we present the explicit expressions as well as the numerical comparison of RC to the lepton current both for fixed scattering angle and transferred momentum squared. Such RC include hard real photon emission from the initial and final leptons, vacuum polarization and vertex correction. The presented RC are charge-even,  therefore they directly apply to a sum of positron- and electron scattering cross sections that could be measured by combined experiments with added positron capabilities at JLab.

For extraction and cancellation of the infrared divergence we use the covariant approach of Bardin-Shumeiko \cite{BSh}. One of the important advantages of this approach over \cite{MoTsai1969} consists in the independence of the final results from the parameter $ \Delta$.
A similar calculation but for fixed transferred momentum squared was performed in  Ref.~\cite{AGIM2015}.

Among the other recent results on RC calculations to the lepton current with the hard photon emission and keeping the lepton mass, we specifically mention two papers. The first one is by Bucoveanu and Spiesberger \cite{Spiesberger} and includes the second-order RC. The second publication describes FORTRAN code developed by    
Banerjee, Engel, Signer, and Ulrich \cite{Signer} with a calculation of the first order RC to several processes in elastic lepton-lepton and lepton-proton scattering.

The rest of the article is organized as follows. The kinematics of elastic process and radiative process are discussed in detail in Sec.~\ref{kin}. In particular, we show that for the description of hard photon emission at fixed scattering angle the ultrarelativistic approximation is not applicable even for relativistic electron-proton scattering. The hadronic tensor and Born cross section are presented in Sec.~\ref{ht}. The additional virtual particle contributions are given in Sec.~\ref{virt}. For the parameterization of the infrared and ultraviolet divergences the dimensional regularization is used. In the next two sections the real photon emission contribution for both fixed $Q^2$ and fixed scattering
angle is presented. For both cases the infrared divergence is extracted and cancelled using the Bardin-Shumeiko approach \cite{BSh}. The comparative numerical analysis for MUSE \cite{MUSE} and \cite{Jlab,Jlab1} experiments can be found in Sec.~\ref{numres}.  A brief discussion and conclusions are presented in the last section. The details of the approach for the infrared divergence extraction are given in \ref{deltash}. The derivation of the compact expression for the Bardin-Shumeiko function $S_\phi$
can be found  in \ref{sfgen}. 
  
\section{Elastic and inelastic processes}
\label{kin}
The unpolarized elastic $lp$-scattering
\be
l(k_1)+p(p_1) \to l^\prime(k_2)+p'(p_2),
\ee
is considered first. Here $k_1$ and $p_1$ ($k_2$ and $p_2$) are the four-momenta of the initial (final) lepton and proton respectively ($k_1^2=k_2^2=m^2$, $p_1^2=p_2^2=M^2$). Although we consider this process in the target rest frame (${\bf p_1}=0$), after definition of the virtual photon momentum as $q=k_1-k_2$,
it will be useful to introduce the kinematic invariants:
\be
&\displaystyle
S=2p_1k_1,\; Q^2=-q^2,\; X=S-Q^2,
\nonumber\\[1mm]
&\displaystyle
\lambda_S=S^2-4m^2M^2,\;
\lambda_X=X^2-4m^2M^2,
\nn [1mm]&\displaystyle
\l_m=Q^2(Q^2+4m^2),
\ee
in such a way, that the energies  of the initial ($k_{10}$) and final ($k_{20}$) leptons as well as the absolute value of their three-momenta ($|{\bf k_1}|$ and $|{\bf k_2}|$, respectively) read: 
\begin{eqnarray}
k_{10}=\frac S{2M},\;
|{\bf k_{1}}|=\frac {\sqrt{\lambda_S}}{2M},\;
%nonumber\\[2mm]
k_{20}=\frac X{2M},\;
|{\bf k_{2}}|=\frac {\sqrt{\lambda_X}}{2M}.\;
%\nonumber\\
\end{eqnarray}

In the present paper we will consider two types of the cross sections: $d\sigma/dQ^2$ and $d\sigma/d\cos\theta$ where the cosine of the scattering angle $\theta$ can be expressed through the invariants:
\be
\cos\theta=\frac {{\bf k_1}\cdot {\bf k_2}}{|{\bf k_1}||{\bf k_2}|}
=\frac{SX-2M^2(Q^2+2m^2)}{\sqrt{\lambda_S\lambda_X}}.
\label{cth}
\ee

Taking into account $X=S-Q^2$, the quadratic equation over $Q^2$ has two solutions
\be
Q^2_\pm=\l_S\frac{S\sin^2\theta +2M^2\pm2M\cos\theta\sqrt{M^2-m^2\sin^2\theta  }}{(S+2M^2)^2- \l_S\cos^2\theta},
\nonumber\\
\label{q2pm}
\ee
where the direct substitution into (\ref{cth}) shows
that $Q^2_-$ is the correct expression
while $Q^2_+$ corresponds to the scattering on  $180^o-\theta$ angle:
\be
\frac{S(S-Q^2_\pm)-2M^2(Q^2_\pm+2m^2)}{\sqrt{\lambda_S((S-Q^2_\pm)^2-4m^2M^2)}}=\mp\cos\theta.
\ee
The restrictions on the scattering angle $-1<\cos\theta <1$
translate into the kinematical limits for $Q^2$:
\be
0<Q^2<\frac{\lambda_S}{S+m^2+M^2}.
\ee

For the description of the inelastic process caused by real photon emission
\be
l(k_1)+p(p_1)\to l^\prime(k_2)+p^\prime(p_2)+\g(k) 
\label{rad}
\ee
($k^2=0$) three additional variables have to be introduced. We choose the standard set \cite{MASCARAD} of them: inelasticity $v=(p_1+k_1-k_2)^2-M^2$, 
$\tau=kq/kp_1 $ and the azimuthal angle $\phi_k$ between
(${\bf k}_1$,${\bf k}_2$) and (${\bf k}$,${\bf q}$) planes in the rest frame (${\bf p}_1=0$).

Using this set of variables, it is straightforward to show that for real photon emission the expressions for the energy and the three-momentum of the scattering lepton have to be modified:
\be
k_{20}=\frac {X-v}{2M},\;
|{\bf k_{2}}|=\frac {\sqrt{(X-v)^2-4m^2M^2}}{2M}.
\label{k20q2}
\ee
%while the photon energy in the target rest frame can be express through the other invariant $R=2k_1p_1$ as:
%\be
%k_{0}=\frac {R}{2M}=\frac {v}{2M(1+\tau)}.
%\ee

As a result, $\cos \theta$ can be expressed through the inelasticity value
and $Q^2$ in a following way:
\begin{eqnarray}
\cos\theta_R=\frac{S(X-v)-2M^2(Q^2+2m^2)}
{\sqrt{\lambda_S((X-v)^2-4M^2m^2)}},
\label{cthv}
\end{eqnarray}
where we introduce the index $R$ to emphasize that at a fixed $Q^2$ the value of $\cos \theta$  depends on the inelasticity of the radiative process.
The restrictions on the scattering angle $-1<\cos\theta_R<1$ set the upper limit for $v$ at fixed
$Q^2$:
\begin{eqnarray}
v_q=\frac {\sqrt{\lambda_S}\sqrt{\l_m}-Q^2(S+2m^2)}{2m^2}.
%\nonumber\\
\label{vmq2}
\end{eqnarray}

Similar to the non-radiative process, there are two possible ways to express $Q^2$ from Eq.~(\ref{cthv}).  After substitution of the obtained expressions for $Q^2$ into the r.h.s. of Eq.~(\ref{cthv}), the correct solution here is:
\begin{eqnarray}
Q_R^2(v)&=&\frac 1{(S+2M^2)^2-\lambda_S\cos^2\theta}
\nn&&\times
\biggl[(S+2M^2)(\lambda_S-vS)
-\lambda_S (S-v)\cos^2\theta
\nn&&
-2 M\sqrt{\lambda_S}
\sqrt{{\cal D}}\cos\theta\biggl],
\label{q2v}
\end{eqnarray}
where the index $R$ poses the same meaning as in Eq.~(\ref{cthv}), namely, at a fixed $\cos \theta$ the value of $Q^2$ depends on the inelasticity of the radiative process. The quantity
\begin{eqnarray}
{\cal D}=M^2(\lambda_S+v(v-2S))-m^2(\lambda_S\sin^2\theta+4vM^2)
\nonumber\\
\end{eqnarray}
must be positive.  It turns out that the upper limit of $v$ for a given
scattering angle follows from that restriction:
\begin{eqnarray}
v_{\theta}=S+2m^2-
\frac m M \sqrt{(S+2M^2)^2-\lambda_S\cos^2\theta}.\;
\label{vth}
\end{eqnarray}

Notice that minimizing ${\cal D}$ maximizes $Q^2_R$
if $\cos \theta > 0$ and minimizes $Q^2_R$ if $\cos \theta < 0$.
The energy and momentum of the scattering lepton for fixed angle read:
\begin{eqnarray}
k_{20}&=&\frac {S-Q^2_R(v)-v}{2M},\;
\nonumber\\
|{\bf k_{2}}|&=&\frac {\lambda_S-v S-Q^2_R(v)(S+2M^2)}{2M\cos\theta\sqrt{\lambda_S}}.
\label{k20th}
\end{eqnarray}

\begin{figure}
%\vspace*{-5mm}
\includegraphics[width=9cm,height=8cm]{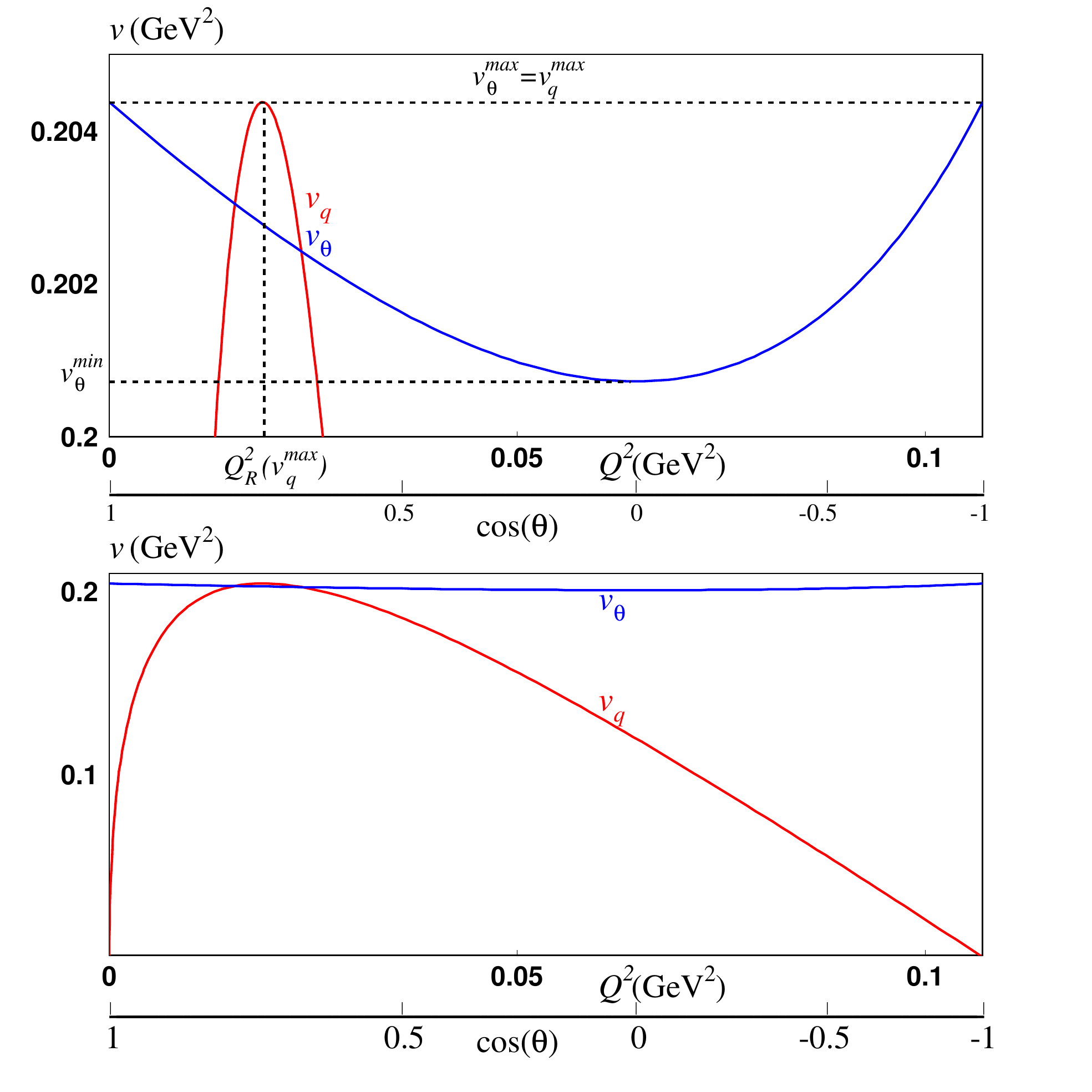}
%\put(-190,-5){\mbox{a)}}
\caption{
The dependence of the upper inelasticity limits $v_q$ and $v_\theta$ on the observable variables 
for the muon beam momentum $|{\bf k}_1|=200$ MeV. Lowest: a full kinematic
range. Upper: a close-up of the region near the kinematic boundary. The quantities $v^{max}_\theta=v^{max}_q$, $v^{min}_\theta $ and $Q^2_R(v^{max}_q)$ are defined by Eqs.~(\ref{vqm}), (\ref{vtm}) and (\ref{q2vmax}) respectively.}
\label{fig2a}
\end{figure}

From Fig.~\ref{fig2a} one can see that when the observable quantity $Q^2$ is close to its kinematical boundaries, the allowed range of the inelasticity reduces to zero that makes it impossible to emit any real photon. The maximum value of the inelasticity 
\begin{eqnarray}
v_q^{max}=S-2m(\sqrt{S+m^2+M^2}-m)
\label{vqm}
\end{eqnarray}
comes at the point that can be obtained after substitution (\ref{vqm}) into (\ref{q2v}), 
\begin{eqnarray}
Q^2_R(v_q^{max})=\frac{m(S+2m^2)}{\sqrt{S+m^2+M^2}}-2m^2.
\label{q2vmax}
\end{eqnarray}

From the upper plot of Fig.~\ref{fig2a} we can see that for fixed angle the upper inelasticity limit reaches its maximum value $v_{\theta}^{max}=v_q^{max}$ at the kinematical boundaries $\cos \theta=\pm 1$ and has a minimum  
\begin{eqnarray}
v_{\theta}^{min}=(S-2mM)\left(1-\frac mM\right)
\label{vtm}
\end{eqnarray}
at $\cos\theta=0$.

\begin{figure}
\vspace*{-8mm}  
%\hspace*{-5mm}  
\includegraphics[width=84mm,height=84mm]{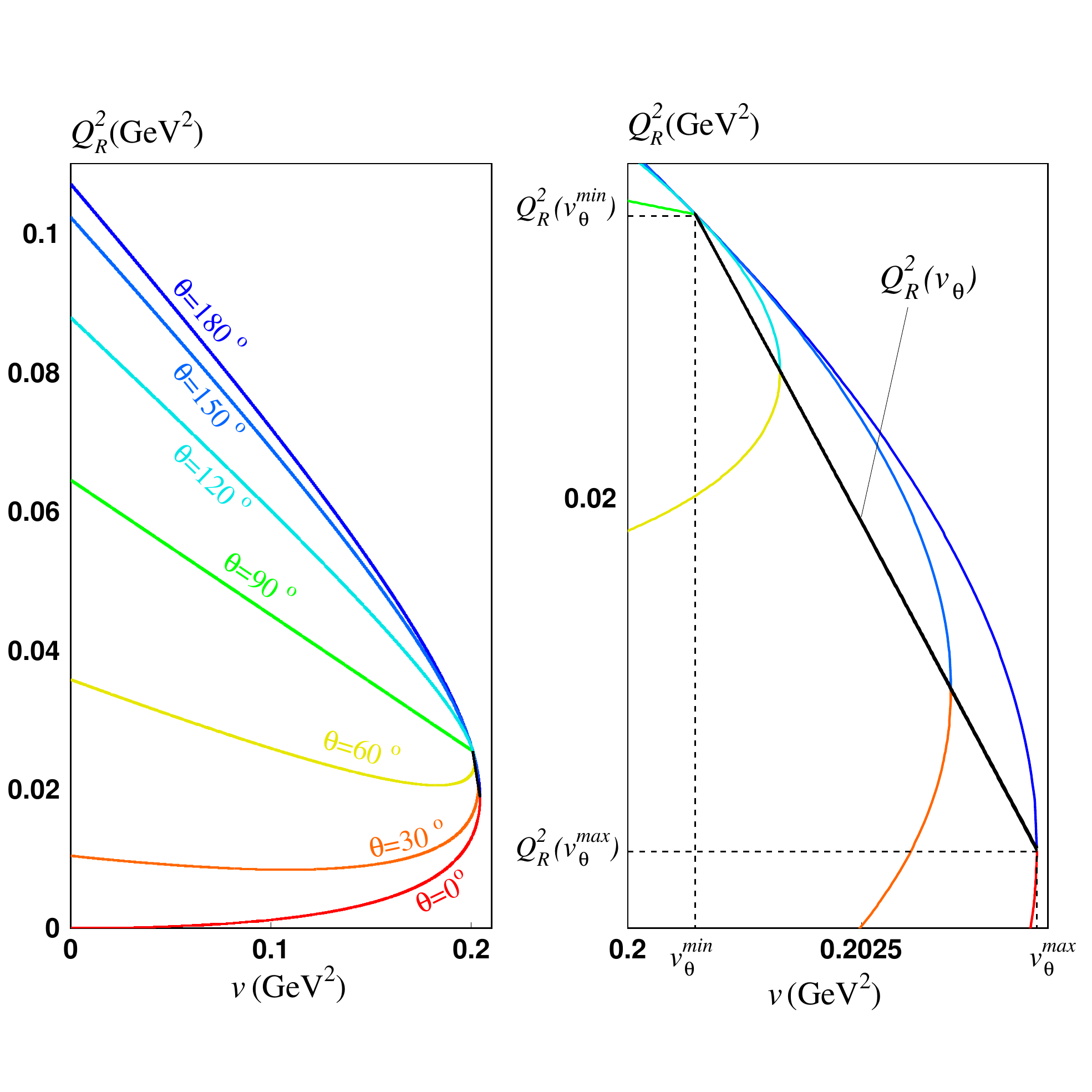}
%\put(-190,-5){\mbox{a)}}
\vspace*{-8mm}  
\caption{ The dependence of $Q^2_R$ on $v$, as given by Eq. (\ref{q2v}) for the different 
fixed scattering angles and the muon beam momentum $|{\bf k}_1|=200$ MeV. Left: a full kinematic
range. Right: a close-up of the region near the kinematic boundary. The joining curves $\theta$ and $180^o-\theta$ line describes  by Eq.~(\ref{q2rvt}). The two indicated points are ($v_\theta ^{min}$, $Q^2_R(v_\theta ^{min})$) according to Eqs.~(\ref{vtm}) and  (\ref{q2rvtm}), and ($v_\theta ^{max}$, $Q^2_R(v_\theta ^{max})$) according to Eqs.~(\ref{vqm}) and  (\ref{q2vmax}).}
\vspace*{-5mm}  
\label{fig2b}
\end{figure}

The dependence of $Q^2_R$ on the inelasticity at different fixed angles is presented in 
Fig.~\ref{fig2b}. From this plot one can see that even for $\theta=0^o$ real photon emission
is not prohibited by any kinematical restrictions.
Opposite to the elastic process, the scattering under zero angle induces non-zero transferred momentum.

After substitution of (\ref{vth}) into (\ref{q2v})
we find the line with boundary common points for $\theta$ and $180^o-\theta$ curves 
\be 
Q^2_R(v_\theta )=\frac{m(S(S+2M^2)-\l_S\cos^2\theta)}{M\sqrt{(S+2M^2)^2-\lambda_S\cos^2\theta}}-2m^2
\label{q2rvt}
\ee
as it is presented in the right plot of Fig.~\ref{fig2b}. The quantity   
$Q^2_R(v_\theta^{max} )$ is defined by Eq.~(\ref{q2vmax}) while
\be
Q^2_R(v_\theta^{min} )=m\Biggl(\frac SM-2m\Biggr).
\label{q2rvtm}
\ee
From Eq.~(\ref{q2rvt})
it can be seen that for the description of  hard photon emission at fixed scattering angle even for high-energy electron-proton scattering the ultrarelativistic approximation is not applicable.

In practice, however, the contribution of the hard real photon emission to the cross section can be essentially reduced by applying a cut $v_{cut}$ on the inelasticity which is also a measured
quantity in the single-arm measurement of the elastically scattered lepton only. Therefore, keeping in mind the inelasticity maximum values, for an upper limit of  this quantity    
both for the fixed $Q^2$ and scattering angle we will use $v_{cut}$ as an experimentally observable variable.

The other invariant quantity $\tau$ can be calculated in the rest frame as
\begin{eqnarray}
\tau=\frac 1M(q_0-|{\bf q}|\cos\theta_k ),
\end{eqnarray}
where $q_0$ (${\bf q}$) is the energy (three-momentum) of the transfer momentum $q$ and $\theta_k$ is the polar angle between the three-momenta ${\bf q}$ and ${\bf k}$. The range of this variable   
is defined through $-1<\cos\theta_k<1 $ and for fixed $Q^2$ and fixed angle $\theta$ it reads:
\begin{eqnarray}
\tau^{q}_{max/min}&=&\frac{Q^2+v\pm\sqrt{\lambda_q}}{2M^2},
\nn
\tau^{\theta }_{max/min}&=&\frac{Q^2_R(v)+v\pm\sqrt{\lambda_v}}{2M^2}
\label{tqt}
\end{eqnarray}
with $\lambda_q=(Q^2+v)^2+4M^2Q^2$ and $\lambda_v=(Q^2_R(v)+v)^2+4M^2Q^2_R(v)$. 

At the end of this Section it is necessary to say about the orientation of the azimuthal photon angle $\phi_k$. It can be defined by choosing a sing in the expression of $\sin \phi_k$ through  the pseudoscalar quantity as 
\be
\sin\phi_k=\pm
\frac{\varepsilon_{\alpha \beta \gamma \delta}p_1^\alpha q^\beta k_1^\gamma k^\delta}{ M |{\bf q}||
{\bf k}_l^\bot| k_0\sin\theta_k },
\ee 
where ${\bf k}_l^\bot$ is the transverse three-momenta
of the incoming or scattering lepton with respect to ${\bf q}$, $k_0$ is a photon energy.
However, during the estimation of the real photon contribution to elastic or inclusive lepton-proton scattering even for polarized particles in contrast to the exclusive or semi-inclusive hadron leptoproduction the sine of $\phi_k$ does not appear for any stage of calculations. Therefore,  we are not concerned about this problem and integrate over $\phi_k$ without taking into account its orientation.

\section{Hadronic tensor and Born contribution}
\label{ht}
\begin{figure}\centering
\scalebox{0.47}{\includegraphics{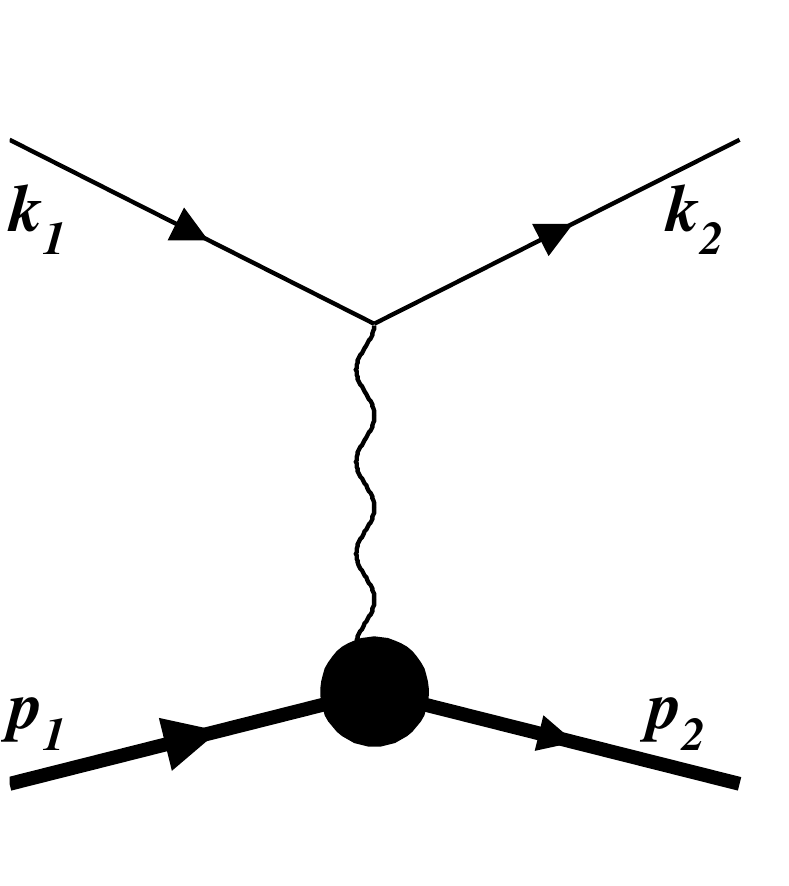}}
\\[-0.1cm]
{\bf \hspace{-.3cm} a)}
\\[0.2cm]
\scalebox{0.47}{\includegraphics{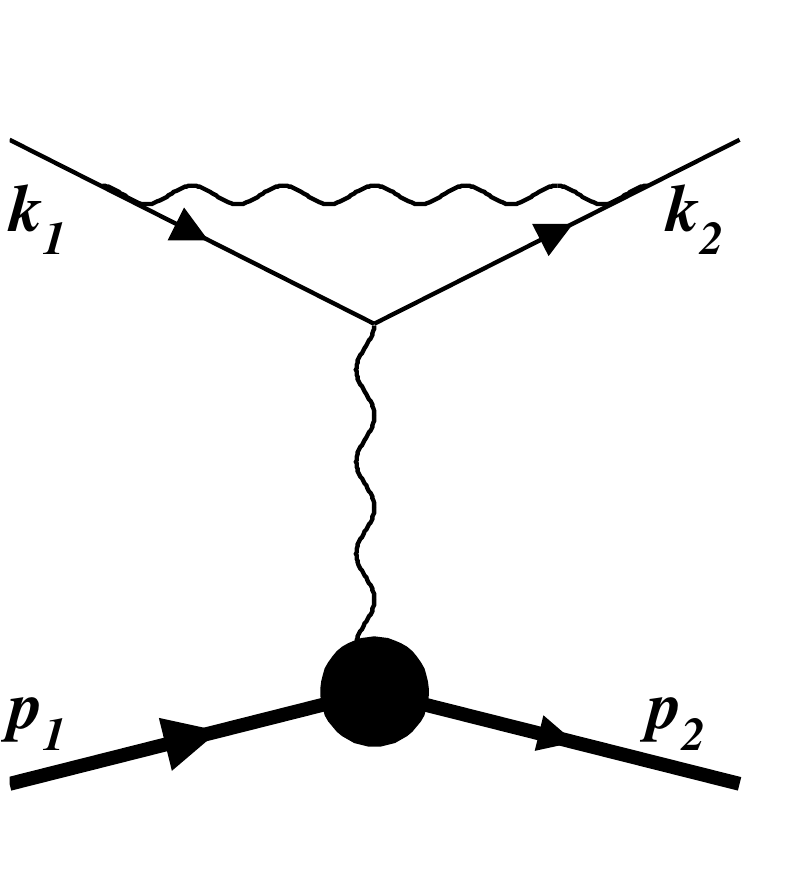}}
\scalebox{0.47}{\includegraphics{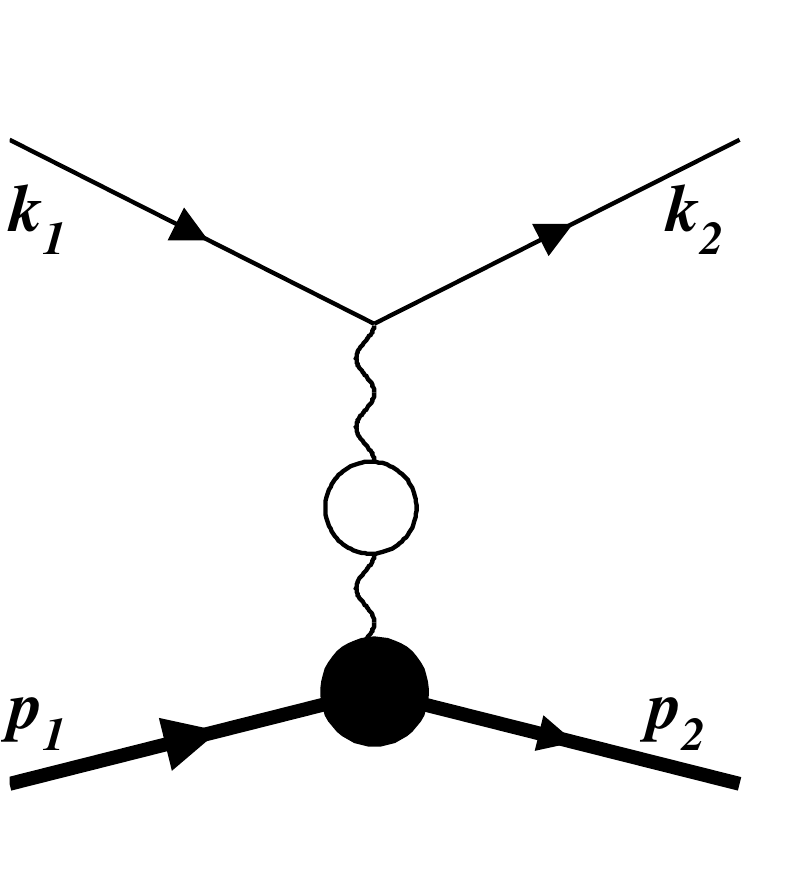}}
\\[-0.1cm]
{\bf \hspace{-.5cm} b) \hspace{3.3cm} c)}
\\[0.1cm]
\scalebox{0.47}{\includegraphics{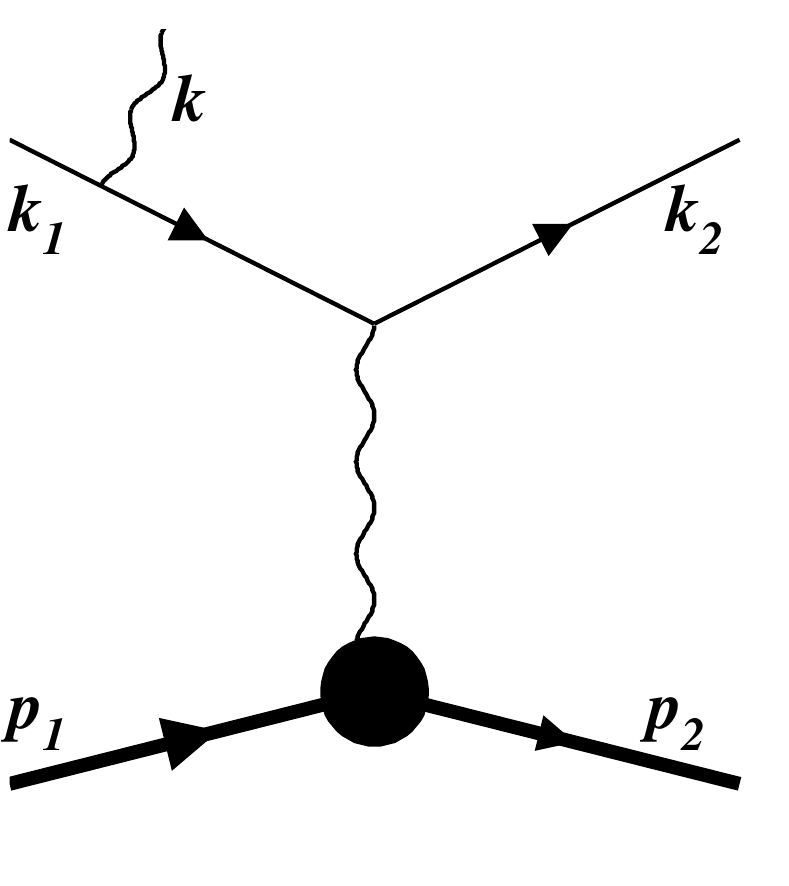}}
\scalebox{0.47}{\includegraphics{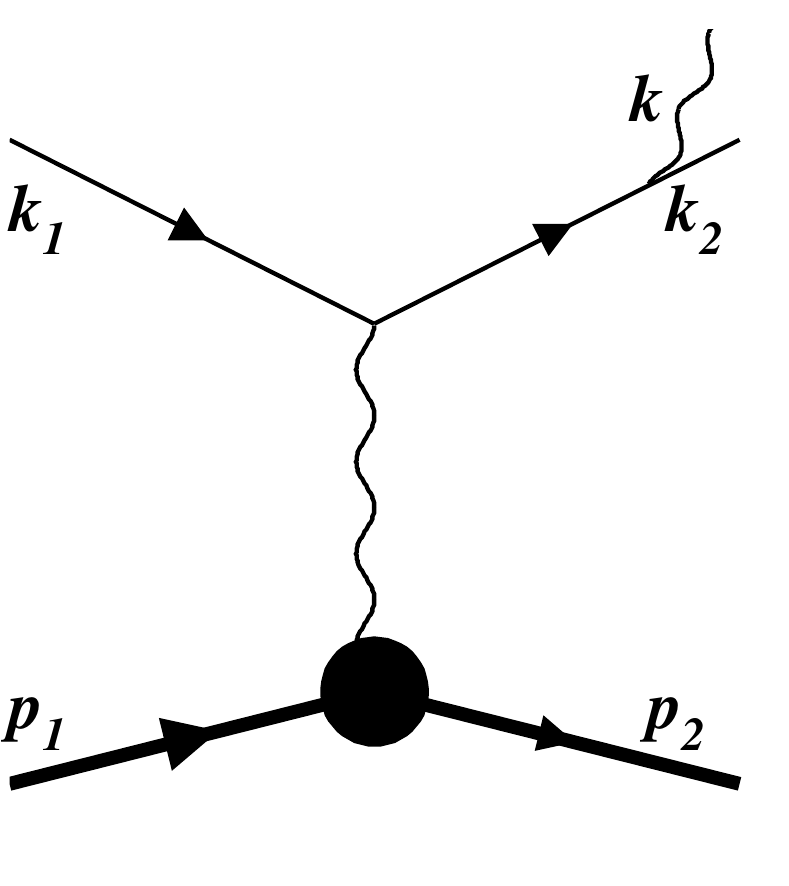}}
\\[-0.1cm]
{\bf \hspace{-.5cm} d) \hspace{3.3cm} e)}
\caption{Feynman graphs corresponding to the Born contribution (a),
leptonic vertex correction (b), vacuum polarization (c), and real photon emission from initial (d) and final (e) leptons.
}
\label{feyn}
\end{figure}

Born contribution to the process depicted by the Feynman graph in Fig.~\ref{feyn}(a)  reads:
\be
d\sigma_B=\frac 1{2\sqrt{\lambda_S}}{\cal M}_B^2d\Gamma_2,
\label{sb1}
\ee
where the phase space has the form
\be
d\Gamma_2&=&\frac 1{(2\pi)^2}\delta^4(p_1+k_1-p_2-k_2)\frac{d^3k_2}{2k_{20}}\frac{d^3p_2}{2p_{20}}
\nn
&=&\frac{dQ^2}{8\pi\sqrt{\lambda_S}}
=\frac{\sqrt{\lambda_X}d\cos \theta}{8\pi(S+2M^2-\cos \theta X\sqrt{\lambda_S/\lambda_X})}.
.
\ee
The matrix element squared is expressed through the convolution of
the leptonic and hadronic tensors
\be
{\cal M}_B^2=\frac {e^4}{Q^4}W_{\mu\nu}(q)L^{\mu\nu}.
\label{mb}
\ee
The leptonic tensor is well known:
\be
L^{\mu\nu}_B&=&\frac 12{\rm Tr}[\g^\mu(\hat k_1+m)\g^\nu(\hat k_2+m)],
\label{lb}
\ee
while the hadronic tensor can be defined through the on-shell proton vertex 
\be
\G_\mu (q)=\g_\mu F_d(-q^2)+\frac{i\sigma_{\mu \nu}q^\nu}{2M}F_p(-q^2),
\ee
where $F_d(F_p)$ is Dirac (Pauli) form factor,
in the following way
\be
W_{\mu\nu}(q)&=&\frac 12{\rm Tr}[\G_\mu (q)(\hat p_1+M)\G_\nu (-q)(\hat p_1+\hat q+M)]
\nn
\ee
and then rearranged into covariant form
\be
W_{\mu\nu}(q)&=&-
\Biggl(g_{\mu\nu}-\frac{q_\mu q_\nu}{q^2}\Biggr){\cal F}_1(-q^2)
\nn&&
+\Biggl(p_{1\mu}+\frac{q_\mu}{2}\Biggr)
\Biggl(p_{1\nu}+\frac{q_\nu}{2}\Biggr)
\frac{{\cal F}_2(-q^2)}{2M^2}
\nn
&=&\sum\limits_{i=1}^2w^i_{\mu\nu}(q){\cal F}_i(-q^2).
\ee
Here
\be
{\cal F}_1(-q^2)&=&-q^2(F_d(-q^2)+F_p(-q^2))^2,\qquad
\nn
{\cal F}_2(-q^2)&=&4M^2F_d(-q^2)^2-q^2F_p(-q^2)^2.
\ee

As a result, after convolution we have
\be
\frac{d\sigma_B}{dQ^2}&=&\frac {2\pi\alpha^2}{\lambda_S Q^4}\sum\limits_{i=1}^2\theta _B^i{\cal F}_i(Q^2),
\nn
\frac{d\sigma_B}{d\cos\theta}&=&j_\theta\frac{d\sigma_B}{dQ^2},
\label{brn}
\ee
where 
\be
j_\theta=\frac{\sqrt{\lambda_S}\lambda_X^{3/2}}{2M^2 (S X-2m^2(Q^2+2M^2))},
\ee
and
\be
\theta _B^1=Q^2-2m^2,\qquad
\theta _B^2=\frac{S X-M^2Q^2}{2M^2}.
\ee
\section{Additional virtual particle contribution}
\label{virt}
The additional virtual particle contribution can be expressed  through Eqs.~(\ref{sb1},\ref{mb})
with replacement of the leptonic tensor (\ref{lb}) by
\begin{eqnarray}
L_{V}^{\mu \nu}&=&
\frac 12 {\rm Tr}[({\hat k}_2+m)\Gamma_{V}^{\mu }({\hat k}_1+m)\gamma^{\nu}]
\nonumber \\&&
+
\frac 12 {\rm Tr}[({\hat k}_2+m)\gamma^{\mu }({\hat k}_1+m){\bar \Gamma}^{\nu}_{V} ],
\end{eqnarray}
where the leptonic vertex $\Gamma_{V}$ contains  the sum of both the lepton vertex correction 
$\Lambda^{\mu }$ and vacuum polarization by lepton $\Pi^{l\mu}_\alpha $ represented by 
the Feynman graphs in Fig.~\ref{feyn}(b) and  Fig.~\ref{feyn}(c), respectively
\begin{eqnarray}
\Gamma_{V}^{\mu }&=&\Lambda^{\mu }+
\Pi^{l\mu}_\alpha 
\gamma^\alpha 
%+\frac \alpha {2\pi } \delta_{\rm vac}^h\gamma^\mu 
,
\nn 
{\bar \Gamma}^{\nu}_{V}&=&\gamma_0\Gamma^{\nu\;\dagger}_{V}\gamma_0.
\end{eqnarray}
Similar to \cite{AGIM2015} we do not consider the vacuum polarization by the hadron.

Since $\Lambda_{\mu }$ and $\Pi^{l\mu} _\alpha $
contain the ultraviolet divergence while
$\Lambda_{\mu }$ also includes the infrared divergent terms,
both of these contributions have to be calculated analytically, and we choose 
dimensional regularization for this calculation.

After the analytical calculation -- detail of which can be found in Appendix D of \cite{HAPRAD3} --
$\Lambda_{\mu }$ and $\Pi^i_{\alpha \mu} $ read:
\begin{eqnarray}
\Lambda_{\mu }&=&\frac {\alpha}{2\pi}
\biggl(\delta_{\rm vert}^{UV}(Q^2)\gamma _\mu-\frac 12 mL_m[{\hat q},\gamma _\mu]\biggr),  
\nonumber \\
\Pi^l_{\alpha \mu }&=&\frac {\alpha}{2\pi}
\biggl(g_{\alpha \mu}+\frac{q_{\alpha}q_{\mu}}{Q^2}\biggr)
\sum_{i=e,\mu,\tau}
\delta_{\rm vac}^{i\; UV}(Q^2).
\label{lp2}
\end{eqnarray}
The term in $\Lambda_{\mu }$ proportional to
\be
L_m=\frac 1\slm\log\frac{\slm+Q^2}{\slm-Q^2}
\label{lm}
\ee
is the anomalous magnetic moment whose contribution reads
\be
\frac{d\sigma_{AMM}}{dQ^2}&=&\frac {\alpha ^3m^2L_m}
{2M^2Q^2\l_S}
\nn &&\times
\Biggl[12M^2{\cal F}_1(Q^2)
%\nn&&\qquad\qquad\qquad
-(Q^2+4M^2){\cal F}_2(Q^2)
\Biggl],
\nn
\frac{d\sigma_{AMM}}{d\cos\theta}&=&j_\theta\frac{d\sigma_{AMM}}{dQ^2}.
\label{amm}
\ee

The ultraviolet divergence contained in the remaining terms of Eqs~(\ref{lp2})
can be removed by applying the mass-shell renormalization procedure   
that requires their vanishing at $Q^2\to 0$:
\be
\delta_{\rm vert}&=&\delta_{\rm vert}^{UV}(Q^2)-\delta_{\rm vert}^{UV}(0),
\nn
\delta_{\rm vac}^i&=&\delta_{\rm vac}^{i\; UV}(Q^2)-\delta_{\rm vac}^{i\; UV}(0).  
\ee
As a result, we obtain
\begin{eqnarray}
\label{dvrtvac}
\delta_{\rm vert}&=&-J_0\biggl(P_{IR}+\log\frac m\mu \biggr)-2  
+\biggl( \frac 32 Q^2+4m^2 \biggr)L_m
\nonumber \\
&&-\frac{Q^2+2m^2}{\sqrt{\lambda_m}}
\biggl(\frac 12\lambda_mL_m^2
+2{\rm Li}_2\biggl[\frac{2\sqrt{\lambda_m}}{Q^2+\sqrt{\lambda_m}} \biggr]
\nonumber \\&&
-\frac{\pi^2}2 \biggr), 
\nonumber \\
\delta_{\rm vac}^l&=&\sum_{i=e,\mu,\tau}\delta_{\rm vac}^i=\sum_{i=e,\mu,\tau}
\Bigl[\frac 23(Q^2+2m^2_i)L_m^i
\nonumber\\&&
-\frac {10} 9
+\frac{8m_i^2}{3Q^2}\Bigl(1-2m^2_iL_m^i
\Bigr)
\Bigr].
\end{eqnarray}
Here
\be
J_0=2((Q^2+2m^2)L_m-1),
\label{j0}
\ee 
$\mu$ is an arbitrary parameter of the dimension of a mass,
\begin{eqnarray}
P_{IR}=\frac 1{n-4}+\frac 12\gamma _E+\log\frac 1{2\sqrt{\pi}}
\label{pir}
\end{eqnarray}
is the infrared divergent term,
\begin{eqnarray}
{\rm Li}_2(x)=-\int\limits^x_0\frac{\log|1-y|}y dy
\end{eqnarray}
is Spence's dilogarithm, and
\be
L_m^i=\frac 1{\sqrt{\l_m^i}}\log\frac{\sqrt{\l_m^i}+Q^2}{\sqrt{\l_m^i}-Q^2},\;
\l_m^i=Q^2(Q^2+4m_i^2).
\label{lmi}
\ee

Finally, the virtual particle contribution reads
\be
\frac{d\sigma_V}{d\zeta}=
\frac{d\sigma_{AMM}}{d\zeta}
+\frac \alpha\pi (\delta_{vert}+\delta_{vac}^l)\frac{d\sigma_B}{d\zeta},
\label{svac}
\ee 
where $\zeta=Q^2$ or $\cos\theta $.

It should be noted that the above obtained expressions for the virtual particle contributions agree  with the results given in Section 3 of \cite{Kaiser} and in Appendix A of \cite{Vanderhaeghen}.  
Particularly, while the comparison with \cite{Vanderhaeghen} is straightforward,
to verify agreement of our results with \cite{Kaiser} we present Eq.~(\ref{svac}) in the electron-muon scattering limit: $F_d\to 1$, $F_p\to 0$, $M\to m_\mu$ and $m\to m_e$. 

\section{Real photon emission for fixed $Q^2$}
The contribution of real photon emission from the lepton leg presented in Fig.~\ref{feyn}(d, e) has a form:
\be
d\sigma_R=\frac 1{2\sqrt{\lambda_S}}{\cal M}_R^2d\Gamma_3,
\ee
where the phase space can be expressed through the photonic variables introduced after (\ref{rad})
\be
d\Gamma_3&=&\frac 1{(2\pi)^5}\delta^4(p_1+k_1-p_2-k_2-k)\frac{d^3k}{2k_{0}}\frac{d^3k_2}{2k_{20}}\frac{d^3p_2}{2p_{20}}
\nn 
&=&\frac{dQ^2vdvd\tau d\phi_k}{2^8\pi^4(1+\tau)^2\sqrt{\lambda_S\lambda_q}}.
\ee
The matrix element squared reads
\be
{\cal M}_R^2=\frac {e^6}{t^2}W_{\mu\nu}(q-k)L_R^{\mu\nu},
\ee
where $t=-(q-k)^2=Q^2+\tau R$ and $R=2p_1k=v/(1+\tau)$.
The leptonic tensor reads: 
\be
L_R^{\mu\nu}&=&-\frac 12{\rm Tr}[\G_R^{\mu\alpha}(\hat k_1+m)\bar \G_{R\alpha}^\nu(\hat k_2+m)],
\ee
with
\be
\G_R^{\mu\alpha}&=&\Biggl(\frac{k_1^\alpha}{k k_1}-\frac{k_2^\alpha}{k k_2}\Biggr)\g^\mu
-\frac{\g^\mu \hat k\g^\alpha}{2kk_1}
-\frac{\g^\alpha \hat k \g^\mu }{2kk_2},
\nonumber\\
\G_{R\alpha}^{\nu}&=&\Biggl(\frac{k_{1\alpha}}{k k_1}-\frac{k_{2\alpha}}{k k_2}\Biggr)\g^\nu
-\frac{\g^\nu \hat k\g_\alpha}{2kk_2}
-\frac{\g_\alpha \hat k \g^\nu }{2kk_1}.
\ee
It is convenient to introduce the following convolutions integrated over $\phi_k$:
\be
\int\limits_0^{2\pi}d\phi_k
L_R^{\mu\nu}w^i_{\mu\nu}(q-k)&=&-4\pi\sqrt{\lambda_q}\sum\limits_{j=1}^{k_i}R^{j-3}
\nn&& \qquad \; \times
\theta_{ij}(v,\tau, Q^2).
\label{ij}
\ee
Here
$k_i=\{3,4\}$,
$\theta_{i1}(v,\tau, Q^2)=4\theta^i_B F_{IR}$ and the other components of $\theta_{ij}(v,\tau, Q^2)$ tensor read:
\be
\theta_{12}&=&4\tau F_{IR},
\nn
\theta_{13}&=&-4F-2\tau^2 F_d,
\nn
\theta_{22}&=&\frac 1{2M^2}\Biggl[2(Q^2-2\tau M^2-2(1+\tau)S)F_{IR}
\nn &&
+S_p(Q^2F_{1+}+2F_{2-}-\tau S_pF_d)\Biggr],
\nn
\theta_{23}&=&\frac 1{2M^2}\Biggl[(\tau (2\tau M^2-Q^2)+4m^2)F_d-S_pF_{1+}
\nn &&
+2(1+\tau )(\tau S_pF_d+XF_{1+}
+F_{IR}-F_{2-})\Biggr]+2F,
\nn
\theta_{24}&=&-\frac 1{2M^2}\tau (1+\tau )(F_{1+}+(\tau+2) F_d),
\label{theta}
\ee
where $S_p=S+X=2S-Q^2$, and
\be
F_{d}&=&\frac 1\tau \biggl(\frac 1{\sqrt{C_2}}-\frac 1{\sqrt{C_1}}\biggr),\
\nn
F_{1+}&=&\frac 1{\sqrt{C_1}}+\frac 1{\sqrt{C_2}},\;
\nn
F_{2\pm}&=&m^2\biggl(\frac {B_2}{C_2^{3/2}}\pm\frac {B_1}{C_1^{3/2}}\biggr),\;
\nn
F_{IR}&=&F_{2+}-(Q^2+2m^2)F_d,\; 
\nn
F&=&\frac 1{\sqrt{\lambda_q}}.
\ee
Here:
\be
C_1&=&4m^2(Q^2+\tau (Q^2+v)-\tau ^2M^2)
\nn&&
+(Q^2+\tau S)^2,
\nn
C_2&=&4m^2(Q^2+\tau (Q^2+v)-\tau ^2M^2)
\nn&&
+(Q^2+\tau (v-X))^2,
\nonumber\\
B_1&=&\tau (S(Q^2+v)+2M^2Q^2)
\nn&&
+Q^2(S_p-v),
\nonumber\\
B_2&=&\tau ((X-v)(Q^2+v)-2M^2Q^2)
\nn&&
+Q^2(S_p-v).
\ee

As a result, we obtain
\be
d\sigma_R=-\frac{\alpha ^3 dQ^2d\tau dv}{2\l_S(1+\tau)t^2}
\sum\limits_{i=1}^2
\sum\limits_{j=1}^{k_i}
{\cal F}_i(t)
R^{j-2}\theta_{ij}(v,\tau, Q^2).
\nn
\ee

A straightforward integration over the photon phase space is
not possible because of infrared divergence coming from the term with $j=1$ in (\ref{ij})
at the point $v=0$ (or $R=0$).  For the consistent extraction
and cancellation of the infrared divergence we use the Bardin-
Shumeiko approach \cite{BSh}. Following this method, the
identical transformation,
\be
d\sigma_R
=d\sigma_R-d\sigma_R^{IR}+d\sigma_R^{IR}
=d\sigma_R^F+d\sigma_R^{IR},
\ee
allows us to split $d\sigma_R$ into the infrared-free $d\sigma_R^F$ and infrared-dependent $d\sigma_R^{IR}$ parts. The last one can be obtained before integration over $\phi_k$ as a term
factorized in front of the Born cross section:
\be
d\sigma_R^{IR}
&=&\frac 1 R\lim_{R\to 0} Rd\sigma_R
=-\frac{\alpha }{\pi^2}d\sigma_B\frac{vdvd\tau d\phi_k}{2(1+\tau)^2\sqrt{\l_q}}{\cal F_{IR}},
\nn
\ee
where
\be
{\cal F_{IR}}&=&\frac 14\biggl(\frac{k_1}{kk_1}-\frac{k_2}{kk_2}\biggr)^2.
\label{fir}
\ee
Note that
\be
F_{IR}&=&\frac{R^2}{2\pi\slq}\int\limits_0^{2\pi}d\phi_k
{\cal F_{IR}}.
\label{fir1}
\ee
The treatment of the infrared divergence by the Bardin-Shumeiko approach
requires to separate $d\sigma_R^{IR}$ into the soft $\d _S$ and hard $\d _H$ parts 
\be
\frac{d\sigma_R^{IR}}{dQ^2}
=\frac\alpha\pi\d _{IR}\frac{d\sigma_B}{dQ^2}
=\frac\alpha\pi(\d _S+\d _H)\frac{d\sigma_B}{dQ^2}
\ee
by introducing of the infinitesimal inelasticity $\bar v$ 
\be
\d_S=-\frac 1\pi \int\limits_0^{\bar v}dv\int\frac{d^3k}{k_0}\delta((p_1+q-k)^2-M^2){\cal F_{IR}},
\nn
\d_H=-\frac 1\pi \int\limits_{\bar v}^{v_{cut}}dv\int\frac{d^3k}{k_0}\delta((p_1+q-k)^2-M^2){\cal F_{IR}}.
\label{dsh}
\ee
This separation allows us to calculate $\d_S$ in the dimensional regularization by
choosing the individual reference systems for each leptonic propagator $1/kk_1$ and  $1/kk_2$, as well as their combination to make them independent of the azimuthal angle $\phi_k$ while the hard part can be calculated in straightforward way without any regularization.

It can be seen from the explicit expressions for $\d_S$ and $\d_H$ -- details of their calculation can be found in \ref{deltash} -- that for $Q^2\to 0$ both of the m tend to zero and their sum
\be
\d_{IR}&=&J_0\biggl[P_{IR}+\log \frac{v_{cut}}{\mu M }\biggr]
+\frac 12 SL_S+\frac 12 XL_X
\nn&&
+S_\phi(k_1,k_2,p_2)
\ee
does not depend on the separated inelasticity $\bar v$ and contains the infrared term $P_{IR}$ as well as a parameter $\mu$ that have to be cancelled against corresponding terms in $\d_{\rm vert}$.

Therefore RC for fixed $Q^2$ read:
\be
\frac{d\sigma_{RC}}{dQ^2}&=&
\frac\alpha \pi(\d_{VR}+\d_{\rm vac}^l)\frac{d\sigma_B}{dQ^2}
+\frac{d\sigma_{AMM}}{dQ^2}
+\frac{d\sigma_F}{dQ^2}.
\label{rcq2}    
\ee
Here the expression for $\d_{\rm vac}^l$ is defined in Eq.~(\ref{dvrtvac}), $\d_{VR}$ is an infrared-free sum $\d_{\rm vert}$ and $\d_{IR}$:
\be
\d_{VR}&=&\d_{IR}+\delta_{\rm vert}=J_0\log \frac{v_{cut}}{m M }
+\frac 12 SL_S+\frac 12 XL_X
\nn&&
+S_\phi(k_1,k_2,p_2)
-2  
+\biggl( \frac 32 Q^2+4m^2 \biggr)L_m
\nonumber \\
&&-\frac{Q^2+2m^2}{\sqrt{\lambda_m}}
\biggl(\frac 12\lambda_mL_m^2
+2{\rm Li}_2\biggl[\frac{2\sqrt{\lambda_m}}{Q^2+\sqrt{\lambda_m}} \biggr]
\nonumber \\&&
-\frac{\pi^2}2 \biggr). 
\ee
The general expression for $S_\phi(k_1,k_2,p_2)$ is reproduced in \ref{sfgen} and for our case
\be
S_\phi(k_1,k_2,p_2)&=&\frac{Q^2+2m^2}{\sqrt{\lambda_m}}\Biggl(\frac 14\lambda_XL_X^2-\frac 14\lambda_SL_S^2
\nn
&+&{\rm Li}_2\biggl [1-\frac {(X+\sqrt{\lambda_X})\rho}{8m^2M^2}\biggr ]
\nn
&+&{\rm Li}_2\biggl [1-\frac {\rho}{2(X+\sqrt{\lambda_X})}\biggr ]
\nn
&-&{\rm Li}_2\biggl [1-\frac {Q^2(S+\sqrt{\lambda_S})\rho}{2M^2(Q^2+\sqrt{\lambda_m})^2}\biggr ]
\nn
&-&{\rm Li}_2\biggl [1-\frac {2m^2Q^2\rho}{(Q^2+\sqrt{\lambda_m})^2(S+\sqrt{\lambda_S})}\biggr ]
\Biggr),
\nn
\ee
where $\rho=(Q^2+\sqrt{\lambda_m})(S_p-\sqrt{\lambda_m})/\sqrt{\lambda_m}$.

The anomalous magnetic moment contribution is represented by Eqs.~(\ref{amm}).
At last, the finite part of the cross section reads:
\be
\frac{d\sigma_F}{dQ^2}&=&-\frac{\alpha ^3}{2\l_S}\int\limits_0^{v_{cut}}dv
\sum\limits_{i=1}^2
\Biggl[
4\frac{J_0\theta_B^i{\cal F}_i(Q^2)}{vQ^4}
\nn
&+&\int\limits_{\tau_{min}^q}^{\tau_{max}^q}\frac{ d\tau }{(1+\tau)t^2}
\sum\limits_{j=1}^{k_i}
{\cal F}_i(t)
R^{j-2}\theta_{ij}(v,\tau, Q^2)
\Biggr],
\ee
where the integration limits over $\tau$ are defined by Eqs.~(\ref{tqt}).   

\section{Real photon emission for fixed scattering angle}
The phase space for this case reads:
\begin{eqnarray}
d\Gamma_3&=&J_\theta(v)
\frac{vdv d\cos\theta  d\tau d\phi_k}{2^8\pi^4(1+\tau)^2\sqrt{\lambda_S \lambda_v}},
\end{eqnarray}
where 
\begin{eqnarray}
J_\theta(v)&=&\frac {\lambda_S-v S-Q^2_R(v)(S+2M^2)}
{(S+2M^2)^2-\lambda_S\cos^2\theta}
\nonumber\\&\times&
\Biggl(\frac{S+2M^2}{\cos\theta}+M\sqrt{\frac{\lambda_S}{\cal D}}(S-v+2m^2)\Biggr),
\end{eqnarray}
and $J_\theta(0)=j_\theta$.

After some algebra similar to the previous section,  we have:
\be
\frac{d\sigma_{RC}}{d\cos\theta}&=&
\frac\alpha \pi(\d_{VR}+\d_{\rm vac}^l)\frac{d\sigma_B}{d\cos\theta}
+\frac{d\sigma_{AMM}}{d\cos\theta}
+\frac{d\sigma_F}{d\cos\theta}
,
\nn
\label{rcth}    
\ee
while the finite part has the following structure:
\be
\frac{d\sigma_F}{d\cos\theta}=-\frac{\alpha ^3}{2\l_S}\int\limits_0^{v_{cut}}dv
\sum\limits_{i=1}^2
\Biggl[
4j_\theta\frac{J_0\theta_B^i{\cal F}_i(Q^2)}{vQ^4}
%\nn&&
+J_\theta(v)
\nn\times
\int\limits_{\tau_{min}^\theta}^{\tau_{max}^\theta}\frac{ d\tau }{(1+\tau)t^2}
\sum\limits_{j=1}^{k_i}
{\cal F}_i(t)
R^{j-2}\theta_{ij}(v,\tau, Q^2_R(v))
\Biggr].
\nn
\ee

\section{Numerical results}
\label{numres}
Here we present the relative RC which is defined as a ratio of RC to the Born cross section
\be
\d_{RC}=\frac{d\sigma_{RC}/d\zeta}{d\sigma_{B}/d\zeta}
\ee
both for fixed $Q^2$ ($\zeta=Q^2$) and the scattering angle ($\zeta=\cos\theta$)
presented in Eqs.~(\ref{rcq2}) and (\ref{rcth}), respectively. Corresponding Born contributions
are defined by Eqs.~(\ref{brn}).
\begin{figure*}[hbt]
\centering
%\vspace*{-18mm}
\includegraphics[width=70mm,height=70mm]{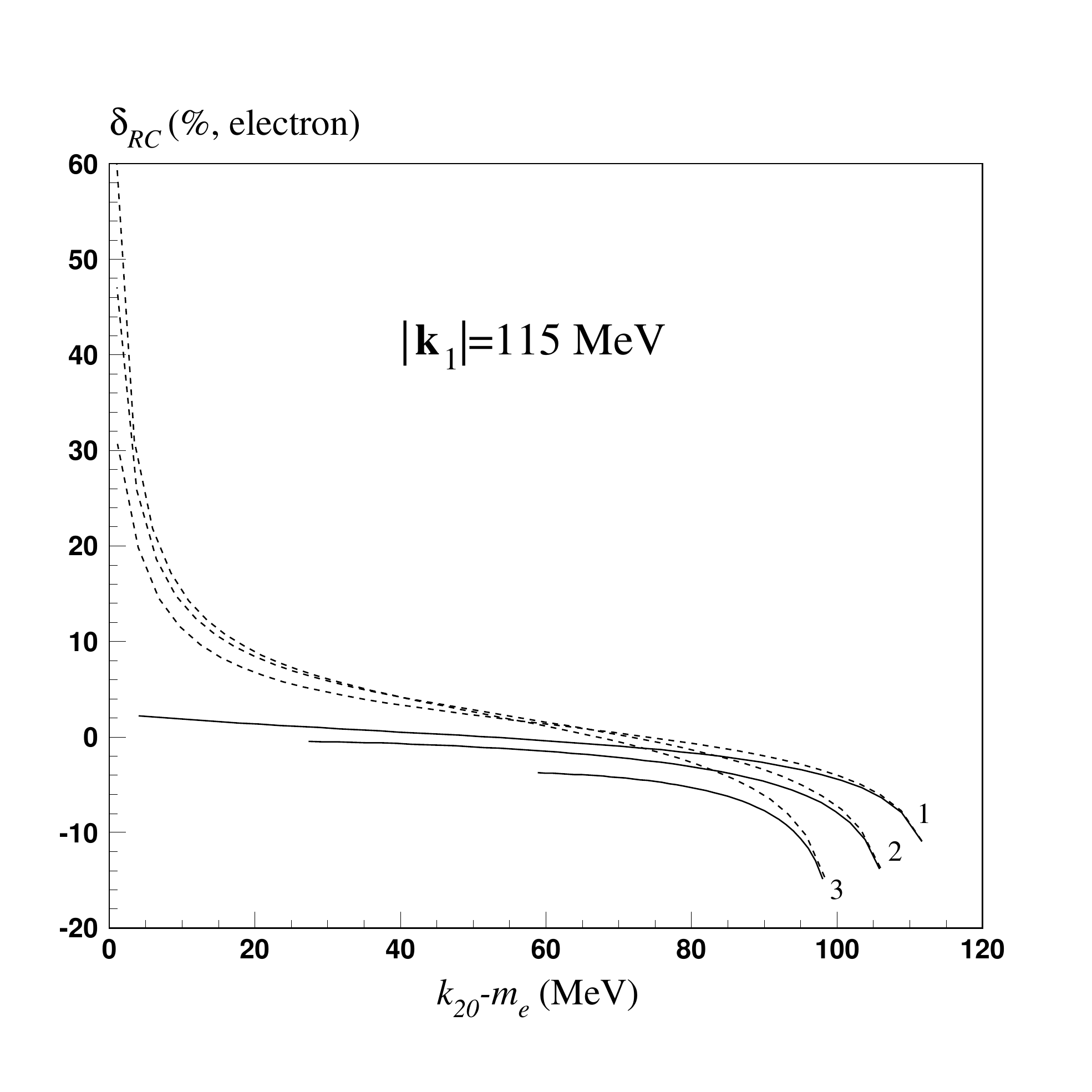}
\hspace*{-6mm}
\includegraphics[width=70mm,height=70mm]{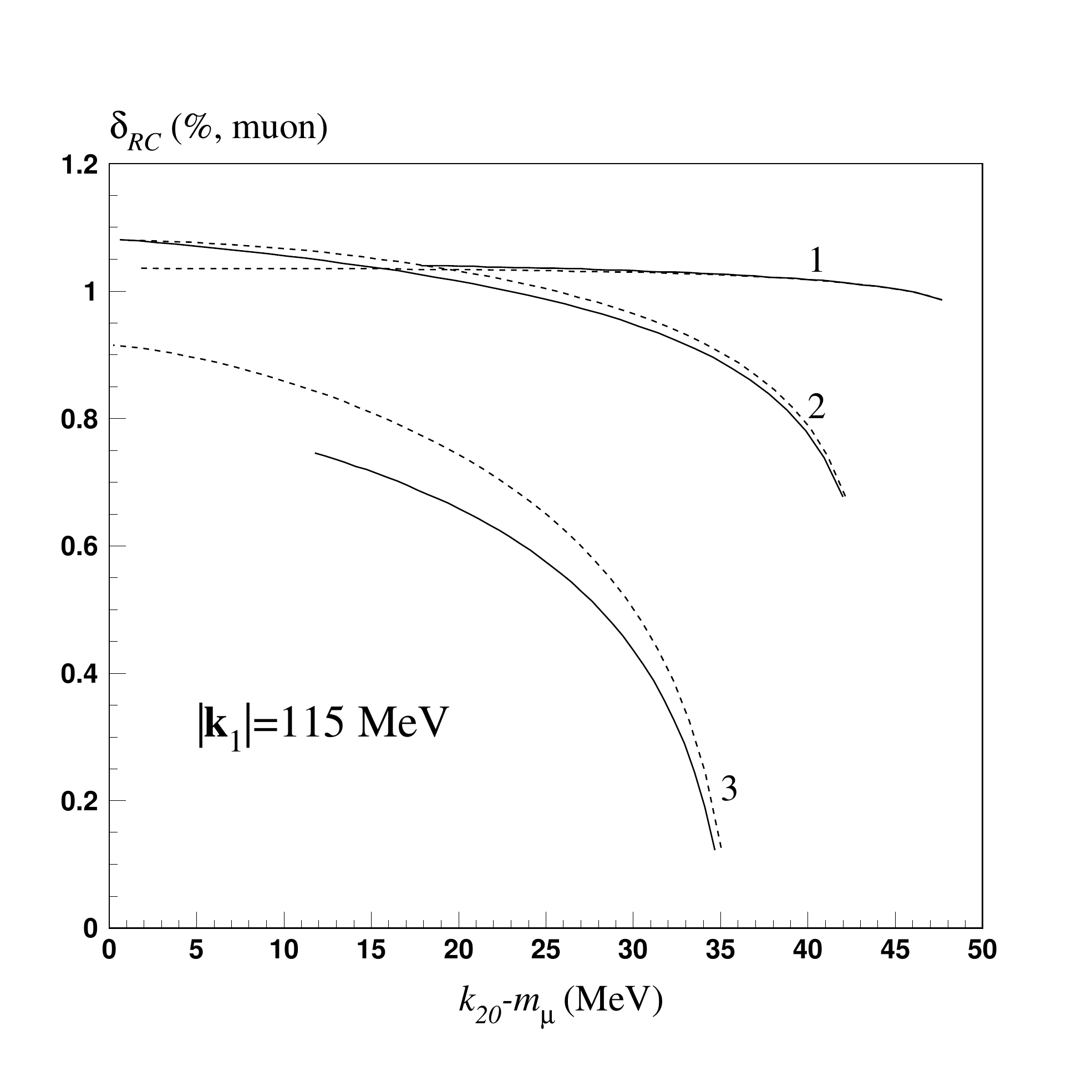}
\\[-9mm]
\includegraphics[width=70mm,height=70mm]{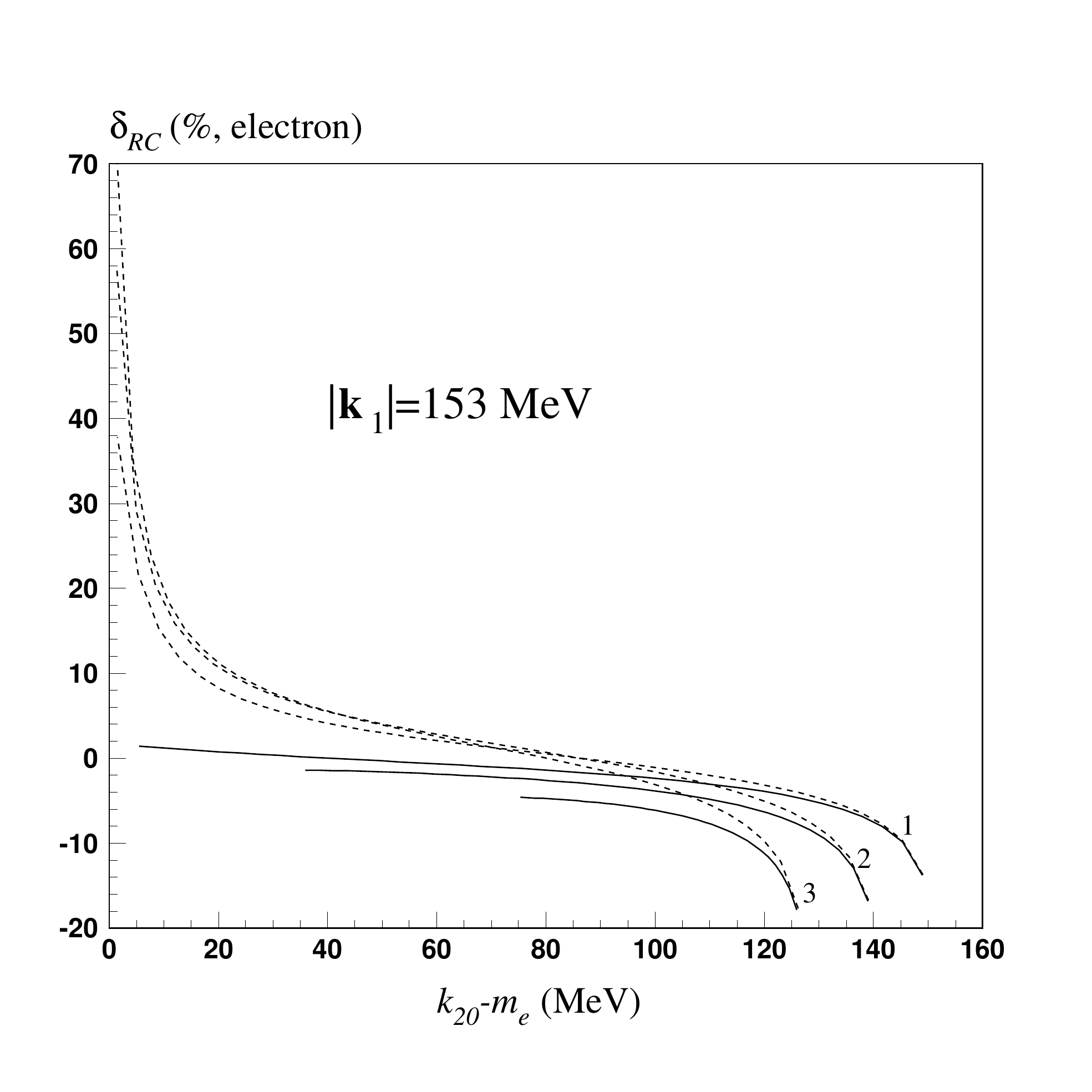}
\hspace*{-6mm}
\includegraphics[width=70mm,height=70mm]{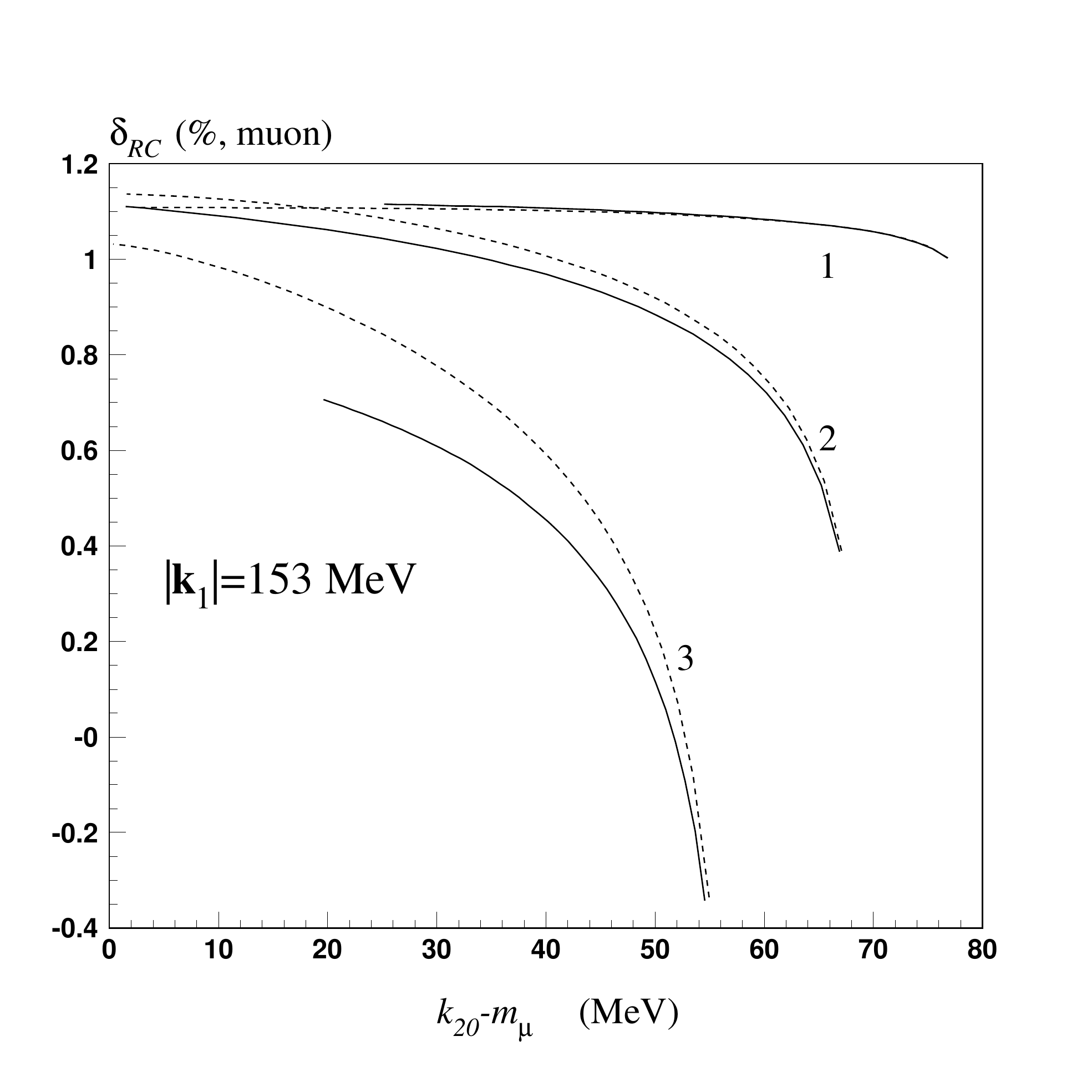}
\\[-9mm]
\includegraphics[width=70mm,height=70mm]{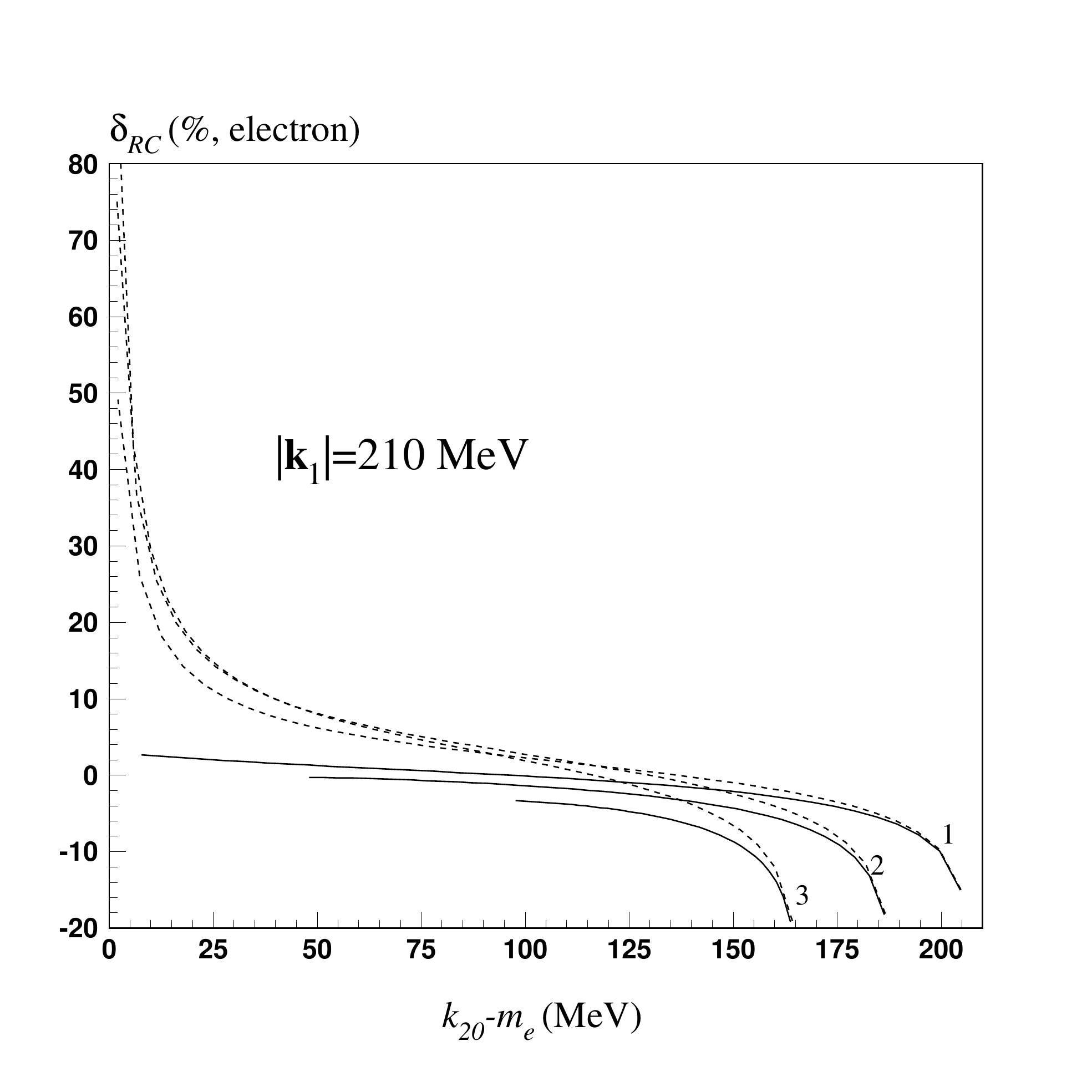}
\hspace*{-6mm}
\includegraphics[width=70mm,height=70mm]{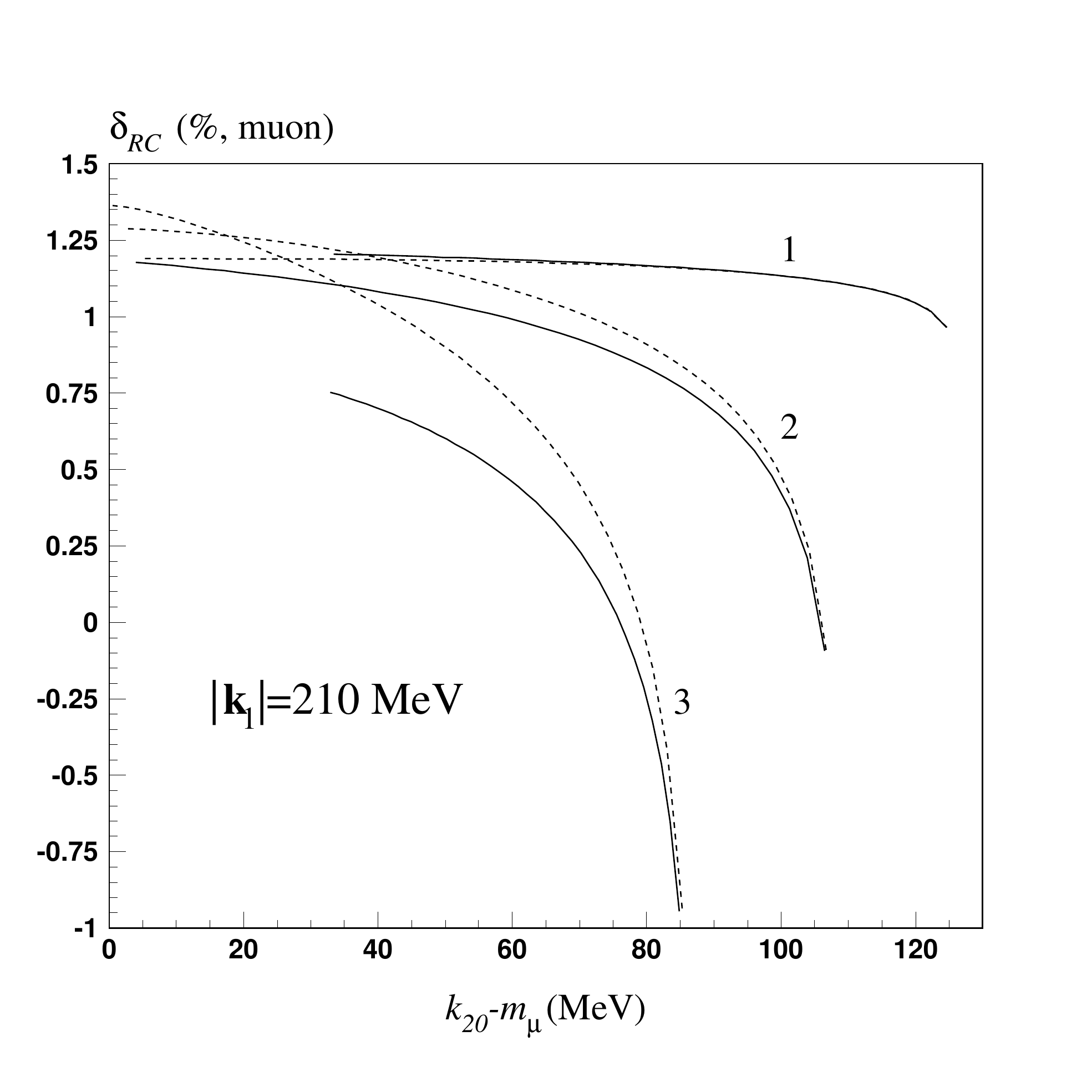}
\\[-4mm]
\caption{
Relative RC vs the value of the scattering lepton kinetic energy for elastic $e p$ and $\mu p$ scattering, beam momenta is equal to 115 MeV, 153 MeV and 210 MeV for $\theta=20^o$ (1), $60^o$ (2), $100^o$ (3). Solid (dashed) line corresponds to fixed $Q^2$ ($\cos\theta $).}
\label{fig3}
\end{figure*}

As mentioned above, a cut applied on the upper  integration limit over inelasticity
allows to reduce the contribution of hard photon emission. On the other hand, for the radiative process the energy of the scattering lepton depends on the inelasticity as it is presented in 
Eqs.~(\ref{k20q2}) for the fixed $Q^2$ and (\ref{k20th}) for the fixed scattering angle. Therefore instead of the upper limit over inelasticity, we can set a cut on the lower limit of the scattered-lepton energy. 

The result of these cuts under MUSE kinematic conditions \cite{MUSE} is presented in Fig.~\ref{fig3}. As we can see, the situation for the scattering electron for fixed $Q^2$ and scattering angle for soft photon emission is almost identical while for hard photon emission it is dramatically different: for the fixed scattering angle RC increase to 80\% while for the fixed $Q^2$ RC do not exceed 5\% .  This is a key observation for both electron and positron scattering in the experimental analysis.

Another interesting issue consists in the $\varepsilon$-behavior at JLab kinematic conditions 
\cite{Jlab,Jlab1}. Following our previous work \cite{chaslet} we can define
\be
\varepsilon_q=\Biggl[1+2\biggl(1+\frac {Q^2}{4M^2} \biggr)\frac{M^2(Q^2-2m^2)}{S(X-v_{cut})-M^2Q^2}\Biggr]^{-1}
\ee
for fixed $Q^2$ and in a similar way
\be
\varepsilon_\theta=\varepsilon_q\big{|}_{Q^2\to Q^2_R(v_{cut})}
\ee
for fixed scattering angle.

\begin{figure}[t]
\centering
\vspace*{-4mm}
%\hspace*{15mm}
\includegraphics[width=75mm,height=75mm]{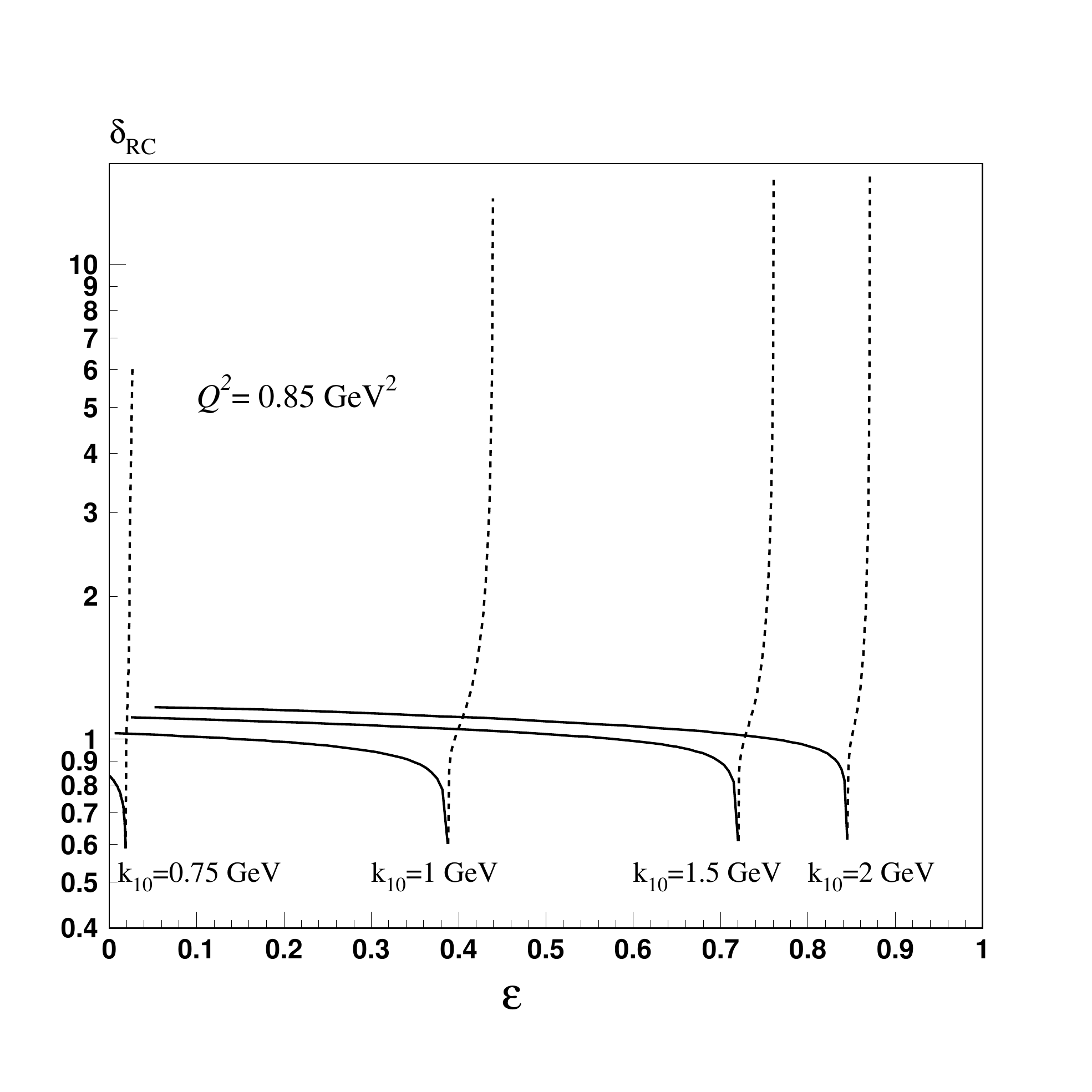}
\\[-9mm]
\includegraphics[width=75mm,height=75mm]{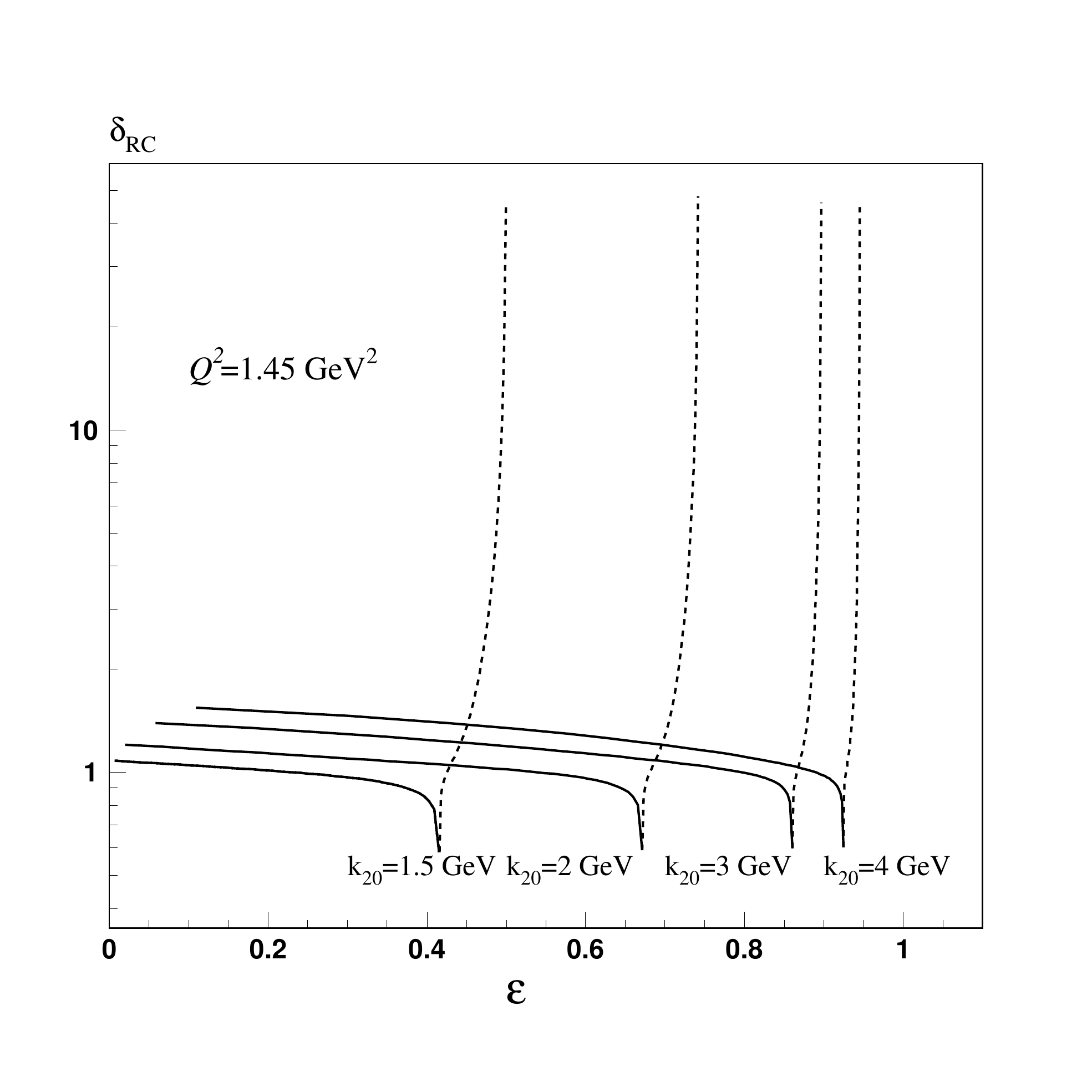}
\caption{
Relative RC vs value of the scattering particle kinetic energy
as a function of $\varepsilon$ at $Q^2=0.85$ GeV$^2$ and $Q^2=1.45$ GeV$^2$.
The solid (dashed) correspond fixed $Q^2$ ($\cos\theta $).
}
\label{fig4}
\end{figure}

The numerical result presented in Fig.~\ref{fig4} shows almost identical
values of RC for the soft photon emission and  different behavior of RC with the hard real photons for the fixed $Q^2$  vs. a fixed scattering angle. In the first case with growing $v_{cut}$ the value of the variable $\varepsilon $ decreases and RC for the hard photon (when $v_{cut}=v_q$) does not exceed 1.6 times the Born contribution, while for the fixed scattering angle $\varepsilon $ goes a little bit up but the absolute value of the relative RC rapidly increases reaching the values up to 45 times (when $v_{cut}=v_\theta$).
Such a rapid change of RC near the kinematic limit of fixed-angle measurements sets more stringent requirements on energy resolution for lepton detection in the fixed-angle kinematic setting, as opposed to fixed $Q^2$ analysis.

\section{Conclusion}
We discussed essential differences between the kinematic description of radiative effects for fixed $Q^2$ vs. fixed scattering angle in the elastic lepton-proton scattering.
In particular, it was shown that for the description of hard-photon emission at fixed scattering angle, even for the high-energy electron-proton scattering the ultrarelativistic approximation is not applicable in the considered kinematics as we approach the limits of phase space. 
The technique of Bardin-Shumeiko for the covariant extraction and cancellation of the infrared divergence as well as the explicit expressions for RC to the lepton current in unpolarized elastic $lp$-scattering within these two cases were presented. The numerical analysis within kinematic conditions of Jefferson Lab measurements and MUSE experiment in PSI has shown the almost identical
values of RC for the soft photon emission and significantly different behavior of RC with the hard real photon for the fixed $Q^2$  compared with fixing a lepton scattering angle. 
The presented formalism may be of use also for the high-energy muon scattering case of the AMBER
proposal \cite{AMBER}.

Based on our recent work \cite{chaslet}, in the nearest future we intend to generalize the numerical comparison of RC calculation for the fixed $Q^2$  and scattering angle with an electron/positron and muon/antimuon charge asymmetry.
We also intend to include simulations of radiative events for the fixed scattering angle into Monte Carlo generator ELRADGEN \cite{ELRADGEN1,ELRADGEN2}, that is used for the hard photon generation in  the elastic $lp$-scattering.

\acknowledgement 
The authors thank the anonymous referee of this paper for insightful comments.

\vspace*{2mm}
\noindent
{\bf Data Availability Statement} This manuscript has no associated data
or the data will not be deposited. [Authors’ comment: The discussion
presented in this article develops from already existing and published
data which are duly referenced.]

\appendix
\section{Calculation of $\delta_S$ and $\delta_H$}
\label{deltash}
For calculation of $\d_S$ in the dimensional regularization
\be
\frac{d^3k^\prime }{k^\prime_0}&\to& \frac{d^{n-1}k^\prime}{(2\pi \mu)^{n-4}k_0}
\nonumber\\
&=&
\frac{2\pi ^{n/2-1}k_0^{\prime n-3}dk_0(1-x^2)^{n/2-2} dx }{(2\pi \mu)^{n-4}\Gamma(n/2-1)},
\label{dnk}
\ee
where $x=\cos \theta $ ($\theta$ is defined as the spatial angle between the photon three-momentum and ${\bf k}^\prime_i $ ($i=1-3$) 
that are introduced below) and $\mu$  is an arbitrary parameter of the dimension
of a mass the reference system ${\bf p}_1{\bf +q=0}$ is used. 

The Feynman parameterization of (\ref{fir}) gives
\be
{\cal F_{IR}}
&=&\frac 1{4k_0^{\prime 2}}\int\limits_0^1dy 
\Biggl[
\frac{m^2}{k_{10}^{\prime 2}(1-x\beta_1)^2}
+\frac{m^2}{k_{20}^{\prime 2}(1-x\beta_2)^2}
\nn &&
-\frac{Q^2+2m^2}{k_{30}^{\prime 2}(1-x\beta_3)^2}
\Biggr]
=\frac 1{4k_0^{\prime 2}}\int\limits_0^1dy {\cal F}(x,y).
\label{ff}
\ee
Here $\beta_i=|{\bf k^\prime}_i|/k^\prime_{i0}$ for $i=1,2,3$ and $k_3=y k_1+(1-y)k_2$.

After the substitution of Eqs.~(\ref{dnk}) and (\ref{ff}) into the definition of $\d_S$ by Eq.~(\ref{dsh}) and, using $\d$-function, integrated over the photon energy $k_0$ 
one can find that
\be
\d_S&=&-\frac 1{2(4\mu\sqrt{\pi})^{n-4}\G(n/2-2)}\int\limits_{-1}^1dx(1-x^2)^{n/2-2}
\nn&&
\times\int\limits_0^1dy{\cal F}(x,y)\int\limits_0^{\bar v}\frac{dv}v\left(\frac vM\right)^{n-4}.
\ee 
The integration over $v$ and the expansion of the obtained expression  into the Laurent series around $n=4$ result in
\be
\d_S&=&\d_S^{IR}+\d_S^1,
\ee
where
\begin{eqnarray}
\delta_S^{IR}&=&-\frac 12\biggl[P_{IR}+\log \frac{\bar v}{\mu M }\biggr] 
\int\limits_0^1dy
\int\limits_{-1}^1dx
{\mathcal F}(x,y)
\end{eqnarray}
 and
\begin{eqnarray}
\delta_S^1&=&-\frac 14
\int\limits_0^1dy
\int\limits_{-1}^1dx\log \biggl[\frac 14(1-x^2)\biggr]
{\mathcal F}(x,y).
\end{eqnarray}
Here $P_{IR}$ is the infrared divergent term defined by
Eq.~(\ref{pir}). Taking into account that  $k_3^2=y(1-y)Q^2+m^2$ the integration
over $x$ and $y$ variables in $\delta_S^{IR}$ is simple:
\be
\delta_S^{IR}&=&J_0\biggl[P_{IR}+\log \frac{\bar v}{\mu M }\biggr],
\ee
where $J_0$ is defined by Eq.~(\ref{j0}).

For the calculation of $\delta_S^1$ we note that in the system ${\bf p}_1{\bf +q=0}$
the energies of the initial and scattering lepton through the invariants:
\be
k_{10}^\prime=\frac X{2M^2},\qquad
k_{20}^\prime=\frac S{2M^2}.
\ee

As a result,
\be
\delta_S^1&=&\frac 12 SL_S+\frac 12 XL_X+S_\phi(k_1,k_2,p_2),
\ee
where
\be
L_S=\frac 1{\sqrt{\l_S}}\log\frac {S+\sqrt{\l_S}}{S-\sqrt{\l_S}},
\nn
L_X=\frac 1{\sqrt{\l_X}}\log\frac {X+\sqrt{\l_X}}{X-\sqrt{\l_X}},
\ee
and
\be
S_\phi(k_1,k_2,p_2)&=&\frac {Q^2+2m^2}4\int\limits_{-1}^1dx\int\limits_0^1dy
\frac{\log[(1-x^2)/4]}{k^{\prime 2}_{30}(1-x \beta_3)^2}.
\nn
\label{sf}
\ee 
Notice that the standard expressions for $S_\phi $ are rather cumbersome, see for example Eqs.~(35) and (A.14) of work \cite{AGIM2015}. In \ref{sfgen} we present a more compact analytical expression for this quantity. 

For the calculation of $\d_H$ the straightforward integration is used. 
Taking into account (\ref{fir1}), one can find that
\be
\d_H&=&-\int\limits_{\bar v}^{v_{cut}}\frac{dv}{v}
\int\limits_{\tau_{min}}^{\tau_{max}}d\tau
F_{IR}
=J_0\log\frac {v_{cut}}{\bar v}.
\ee

\section{Calculation of $S_\phi$}
\label{sfgen}
Here we present a general approach suggested by 't~Hooft and Veltman in their work \cite{tHooft} for a compact representation of the $S_\phi$-function introduced by Bardin and Shumeiko in \cite{BSh}.  
Let us consider a real photon with a momentum $k$ and three other time-like four-momenta $a_i$ ($i=1,2,3$)  with masses $m_i^2=a_i^2$. The basic idea consists in Feynman parameterization. Instead of usual approach used in the standard Bardin-Shumeiko technique with two fermionic propagator presented in previous appendix, taken in the system ${\bf a}_3=0$:
\be
\frac 1{a_1k}\frac 1{ a_2k}=\g \frac 1{a_1k} \frac 1{\g a_2k}=\frac \g{k_0^2}\int\limits_0^1\frac{dy}{a_{40}^2(1-x\beta)^2}.
\ee
Here, as in the previous appendix $x=\cos\theta$, a new four-vector $a_4=y a_1+(1-y)\g a_2$, and $\beta=|{\bf a}_4|/a_{40}$. The quantity $\g $ is choosing in such a way, that $(a_1-\g a_2)^2=0$, $i.e.$ $a_1-\g a_2$ is lightlike vector. 

Now introduce the following invariants:
\be
s_1=2a_1a_3,\; \l_1=s_1^2-4m_1^2m_3^2,
\nn
s_2=2a_2a_3,\; \l_2=s_2^2-4m_2^2m_3^2,
\nn
s_3=2a_1a_2,\; \l_3=s_3^2-4m_1^2m_2^2.
\ee
Then equation $(a_1-\g a_2)^2=0$ has the following two solutions:
\be
\g_1=\frac{2m_1^2}{s_3+\sqrt{\l_3}},\qquad
\g_2=\frac{s_3+\sqrt{\l_3}}{2m_2^2},
\ee
and the generalized form of $S_\phi$ looks as (\ref{sf}):
\be
S_\phi=\frac 14\g s_3\int\limits_{0}^1\frac{dy}{a_{40}^2}\int\limits_{-1}^1dx
\frac{\log [(1-x^2)/4]}{(1-x\beta)^2}.
\ee
The first integration over $x$ is straightforward
\be
S_\phi=\frac 12\g s_3\int\limits_{0}^1\frac{dy}{m_4^2\beta}
\log \frac{1-\beta}{1+\beta},
\ee
where $m_4^2=a_4^2=ym_1^2+(1-y)\g^2 m_2^2$. The second integration has to be performed after the standard substitutions, while taking into account that for the first two momenta $a_{i0}=s_i/(2m_3)$.

Finally, we can find that for the general case $S_\phi$ depends on six variables and for $\g=\g_1$ it has the following structure:
\be
S_\phi (a_1,a_2,a_3)&=&\frac {s_3}{\sl3}
\Biggl(
\log^2\frac {s_1+\sqrt{\l_1}}{2m_1m_3}
-\log^2\frac {s_2+\sqrt{\l_2}}{2m_2m_3}
\nn &&
+{\rm Li}_2\biggl[1-\frac{(s_1+\sqrt{\l_1})\rho}{8m_1^2m_3^2}\biggr]
\nn &&
+{\rm Li}_2\biggl[1-\frac{\rho}{2(s_1+\sqrt{\l_1})}\biggr]
\nn &&
-{\rm Li}_2\biggl[1-\frac{(s_2+\sqrt{\l_2})\rho}{4m_3^2(s_3+\sl3)}\biggr]
\nn &&
-{\rm Li}_2\biggl[1-\frac{m_2^2\rho}{(s_2+\sqrt{\l_2})(s_3+\sl3)}\biggr],
\nn
\ee
where $\rho=(2s_1(s_3+\sl3)-4m_1^2s_2)/\sl3 $.

It should be noted that
\be
S_\phi (a_1,a_2,a_3)=
S_\phi (a_2,a_1,a_3)
\ee
The r.h.s. of this equation  corresponds $\g=\g_2$.

In our case  $a_1=k_1$, $a_2=k_2$, $a_3=p_2$, and  $s_1=X$, $s_2=S$, $s_3=Q^2+2m^2$, $m_1=m_2=m$, $m_3=M$.

\end{document}